\documentclass{aa}  

\usepackage{graphicx}
\usepackage{txfonts}

\begin{document} 

\title{Lithium depletion and angular momentum transport in F-type and G-type stars in Galactic open clusters}

   \authorrunning{T. Dumont et al.}
   
   \titlerunning{Lithium depletion and angular momentum transport in F-type and G-type stars in Galactic open clusters}
   

   \author{T. Dumont
          \inst{1,2} \fnmsep\thanks{e-mail: thibaut.dumont@unige.ch},
          C. Charbonnel \inst{1,3},
          A. Palacios \inst{2},
          and S. Borisov \inst{1,4}
          }

   \institute{Department of Astronomy, University of Geneva, Chemin Pegasi
   51, 1290 Versoix, Switzerland
         \and LUPM, Universit\'e de Montpellier, CNRS, Place Eug\`ene Bataillon, 34095 Montpellier, France
         \and IRAP, CNRS UMR 5277 \& Universit\'e de Toulouse, 14 avenue Edouard Belin, 31400 Toulouse, France
         \and Sternberg Astronomical Institute, M.V. Lomonosov Moscow State University, Universitetsky prospect 13, Moscow, 119234, Russia
             }

   \date{(Received; Revised; Accepted)}

 
  \abstract
   {Open clusters provide unambiguous clues to understand the evolution of $\rm ^7Li$ at the surface of low-mass stars and its possible correlation with stellar rotation, which is a challenge for both stellar hydrodynamics and Galactic chemical evolution.} 
   {We aim to quantify the efficiency of the transport processes for both angular momentum and chemicals that are required  to explain simultaneously the observed behaviour of surface $\rm ^7Li$ (and $\rm ^9Be$) and rotation as well as the internal rotation profiles inferred from helio- and asteroseismology in F- and G-type main sequence stars. 
   }
   {We apply the model for the transport of angular momentum and chemicals that we tailored in a previous work for solar-type stars to an extended range of initial masses and metallicities corresponding to F- an G-type stars in a sample of 20 Galactic open clusters. We evaluate its ability to explain the $\rm ^7Li$, $\rm ^9Be$, and rotation periods observations. This model includes atomic diffusion, rotation-induced processes (for which we tested different prescriptions for shear turbulence), penetrative convection with a rotational dependence, parametric viscosity and turbulence, and magnetic braking.
   }
   {Over the entire range of masses, metallicities, and ages explored, we reproduce the evolution of the surface rotation rates and predict, for the first time, the observed anti-correlation between the surface rotation rate and $\rm ^7Li$ depletion as a consequence of the penetrative convection prescription. The $\rm ^7Li$ behaviour and its evolution with time is well reproduced for G-type stars. However, the ability of the model to reproduce the so-called $\rm ^7Li$ dip centred around $\sim$6600~K strongly depends on the adopted prescriptions for shear turbulence. It also requires a stellar mass dependence for the parametric viscosity adopted for the transport of angular momentum, similar to the behaviour predicted for the generation and luminosity of internal gravity waves generated by stellar convective envelopes. Finally, the model predicts internal rotation profiles in good agreement with asteroseismic constraints in main sequence stars.
   }
   {We provide an efficient way to model G-type stars of different ages and metallicities successfully. However, the $\rm ^7Li$ and $\rm ^9Be$ dip constraints urgently call for further hydrodynamical studies to better model turbulence in stars, and for the exploration of physical processes such as tachocline mixing for the transport of chemicals and internal gravity waves for the transport of angular momentum.
  Finally, additional data for the internal rotation and for $\rm ^9Be$ in main sequence low-mass stars are definitively needed.
   }


   \maketitle
   
%

\section{Introduction}
\label{section:introduction}
Understanding the  evolution of $\rm ^7Li$  (hereafter Li) in low-mass stars is one of the main challenges for stellar and Galactic astrophysics. Despite Li being a very scarce element, it is a tracer of Galactic chemical evolution \citep[e.g.][]{1982A&A...115..357S,Matteucci1995,Romano2001,Travaglio2001,Prantzos2012b} and of transport processes that occur in stellar interiors \citep[e.g.][hereafter Paper I, and references therein]{CharbonneauMichaud1990,1996A&A...305..513M,2000A&A...354..943M,2005Sci...309.2189C,2010IAUS..268..365T,Castroetal2016,2021A&A...646A..48D}. An overall picture of the different possibly involved processes is described in the reviews by \citet[][]{2013LNP...865...23M} and \citet[][]{2019ARA&A..57...35A}.

Observations of open clusters have provided numerous Li abundance data for stars of different ages over a large range of masses and metallicities \citep[e.g.][]{1976PASP...88..353B,DuncanJones1983,Cayrel1984,Balachandran1988,Balachandran2011,Soderblom1990,Soderblom1993,1991ApJ...370L..95B,Thorburn1993,GarciaLopez1994,Swensonetal1994,Jones1999,2003ApJ...583..955B,2003ApJ...582..410B,2005A&A...442..615S,2009AJ....138.1171A,2018AJ....156...37A,2012AJ....144..137C,2017AJ....153..128C,2019AJ....158..163D,2020A&A...640L...1R}. It has been clearly evidenced that Li depletion increases with time and is linked to stellar mass \citep[e.g.][for a review]{2000IAUS..198...61D}. At a given age, the Li behaviour as a function of the stellar effective temperature (T$_{\rm eff}$) shows two specific patterns. On the one hand, photospheric Li abundances of G-type stars decrease with decreasing effective temperature (decreasing mass). On the other hand, a group of F-type stars with T$_{\rm eff}$ centred around 6'600~K fall into the so-called Li dip which appears in open clusters older than $\sim$ 200~Myrs \citep[e.g.][]{Wallerstein1965,1986ApJ...302L..49B,1986ApJ...309L..17H,1993AJ....106.1080S,Balachandran1995,2004ApJ...614L..65S,2016ApJ...830...49B}. 

Classical stellar evolution (accounting only for convection as a mixing process in stellar interiors) predicts noticeable Li depletion during the pre-main sequence (PMS) when the base of the convective envelope is deep enough to reach the Li-burning temperature, which happens only for less massive stars (i.e. below $\sim$ 0.9 - 1.0~M$_{\odot}$ at solar metallicity; e.g.  \citealt[][]{Bodenheimer1965,1997ARA&A..35..557P}). However, it does not predict any further surface Li variation until the first dredge-up occurs after the stars leave the main sequence (MS), in striking contrast with the observational evidence of the steepening along the main sequence of the Li-T$_{\rm eff}$ trend of the cool side of the Li dip and of the Li dip itself. 

The key role of rotation was pointed out as the cool edge of the Li dip coinciding with the so-called Kraft rotation break, as observed for instance in the Hyades \citep{1987PASP...99.1067B} and in NGC 752 \citep{1986ApJ...309L..17H}. The Kraft break \citep{1967ApJ...150..551K}
corresponds to the transition stellar mass ($\simeq$ 1.2~M$_{\odot}$ at solar metallicity) where important structural changes occur in main sequence stars. In particular, stars more massive than the transition value have an extremely thin convective envelope, implying weaker or ineffective magnetic braking compared to cooler, less massive stars with a thick convective envelope that can sustain efficient stellar wind magnetic braking \citep[e.g.][]{Schatzman1962,WeberDavis1967,Matt2015,Kawaler1988,2017AJ....153..128C,2019AJ....158..163D}.
The cool side of the dip also corresponds to the transition region where the convective envelope can efficiently generate internal gravity waves that transport angular momentum \citep{2003A&A...405.1025T,2004A&A...418.1051T,2005A&A...440..981T}, and to an internal structure where the stellar core is radiative while hotter MS stars host a convective core.

Different non-standard mixing-processes (beyond convection) have been explored to explain the observed Li features in F-type and G-type stars.\ This includes the following: convective overshooting or penetrative convection \citep[][Paper I]{1999A&A...347..272S,2017ApJ...836..192B,2017ApJ...845L...6B,2018MNRAS.481.4389J}, atomic diffusion \citep{Michaud1986,1993ApJ...416..312R,Turcotte1998}, mass loss \citep{2010ApJ...713.1108G}, planet accretion or migration \citep{2002A&A...386.1039M,2009A&A...494..663C}, tachocline mixing \citep[][Paper I]{1999ApJ...525.1032B}, internal gravity waves \citep[][]{1994A&A...281..421M}, rotation-induced mixing \citep[][]{1990ApJS...74..501P,1992A&A...255..191C,2003A&A...399..603P,Eggenberger2012,2016ApJ...829...32S,2016A&A...587A.105A,2019ApJ...881..103Z}, magnetic processes and instabilities \citep[][]{1993ApJ...417..762C,1996ApJ...466.1078R,2005A&A...440L...9E,2010A&A...519A.116E}, as well as different combinations of some of the above-mentioned processes where the transports of chemicals and angular momentum are intimately coupled \citep[][Paper I]{1996A&A...312.1000R,2005Sci...309.2189C,2005A&A...440..981T,2020A&A...633A..23D,2020A&A...643A.164S}. 

Importantly, Li appears to be only one piece of a bigger puzzle that should also include the constraints from $\rm ^9Be$ (hereafter Be) that burns at a slightly higher temperature than Li \citep[$\sim$ 3.5~MK and 2.5~MK, respectively; e.g.][]{1997ARA&A..35..557P}. Indeed, while the Be dip coincides with the Li dip  \citep[e.g.][]{1998ApJ...498L.147D,2001ApJ...553..754B,2004ApJ...613.1202B,2020ApJ...888...28B}, Be is hardly depleted in the Sun, solar-like stars, and G-type stars \citep[][]{1998Natur.392..791B,2003ApJ...583..955B,2003ApJ...582..410B,2004ApJ...605..864B,2016ApJ...830...49B,2020ApJ...888...28B}. The Be behaviour thus brings additional constraints to the possible origin of the observed behaviour of Li in F- and G-type stars and to the depth of the required mixing process.
Last but not least, the difficulty is to find the actual connection between the transport of chemicals and the transport of angular momentum at play in stars of different spectral types along their evolution, to account simultaneously for the internal rotation profiles that can be deduced for the Sun and for some other stars thanks to asteroseismology \citep[][]{1988SvAL...14..145K,1995Natur.376..669E,2008A&A...484..517M,2008ApJ...679.1636E,2013A&A...549A..74M,2015MNRAS.452.2654B,2019A&A...626L...1E,2019LRSP...16....4G}.

In this work, we explore the chemical and rotational evolution of low-mass stars on the PMS and the MS for different stellar masses and metallicities that cover the range of Galactic open clusters with ages between 5 Myrs and 4 Gyrs. We study the effects and the relevance for these stars of different transport processes that we already explored and validated for the specific case of the solar-type stars (Paper I), and that depend on both mass and metallicity. In Sect.~\ref{sec:obs}, we present the observational data that we used to constrain  model predictions. In Sect.~\ref{STAREVOL} we describe the input physics of the stellar models and the different processes implemented in the evolution code STAREVOL used for this work. In Sect.~\ref{sect:predictions}, we compare the observations for the Hyades and Praesepe open clusters with the predictions for Li, Be, and surface rotation of our non-standard model including meridional circulation, shear-induced turbulence, atomic diffusion, overshoot, and parametric viscosity and turbulence. In Sect.~\ref{sect:multicluster}, we compare the model predictions with Li and rotational periods ($P_{\rm rot}$) data in a sample of open clusters over a broad range in age and metallicity. In Sect.~\ref{sect:Discussion}, we discuss model predictions for the internal rotation and compare to asteroseismic measurements. We summarise our results and conclude in Sect.~\ref{sect:CONCLUSION}. 
   
\section{Observational data}
\label{sec:obs}
We use observational data for a sample of Galactic open clusters. Their names, ages, metallicities, and distances to the Galactic centre are given in Table~\ref{tab:OCinfos1} along with bibliographical references from which the Li and Be surface abundances and rotation periods of individual stars were taken. We only take into account non-binary stars with confirmed membership as mentioned or flagged in the reference papers cited in Table~\ref{tab:OCinfos1}.

\begin{table*}[t!]
    \centering
    \caption{Main properties of the open clusters considered in this work, with the corresponding references for the Li and Be abundances and the rotation periods.}
    \begin{tabular}{c|c|c|c|c|c|c|c}
         Name & Age (Myrs) & Ref Age & [Fe/H] & Distance (kpc) & Ref Li & Ref Be & Ref $P_{\rm rot}$\\
         \hline \hline 
         NGC 2362 & 5 & I & 0.00 & 9.11 & - & - & a \\
         NGC 2547 & 35 & I & -0.14$\pm$0.10 & 8.39 & - & - & b \\
         IC 2602 & 35 & II & -0.02$\pm$0.02 & 8.29 & 1 & - & - \\
         IC 2391 & 36 & II & -0.03$\pm$0.02 & 8.34 & 1 & & c \\
         Pleiades (Melotte 22) & 87 & II & -0.01$\pm$0.05 & 8.45 & 2 & - & b;d \\
         $\alpha$Per (Melotte 20) & 87 & II & +0.14$\pm$0.11 & 8.48 & 3;4 & - & c \\
         M 35 (NGC 2168) & 150 & I & -0.17$\pm$0.01 & 9.24 & 5;6 & - & e \\
         M 50 (NGC 2323) & 150 & I & 0.00$^\bullet$ & 9.10 & - & - & b;f \\
         NGC 2516 & 150 & I & -0.06$\pm$0.05 & 8.32 & - & - & b \\
         M 37 (NGC 2099) & 500 & I & +0.02$\pm$0.05 & 9.77 & - & - & b;g \\
         Coma Ber (Melotte 111) & 570 & III & 0.00$\pm$0.08 & 8.35 & 7 & - & -\\
         UMa & 600 & IV & -0.09$^\star$ & 8.37$^\ast$ & 7 & - & -\\
         Hyades (Melotte 25) & 720 & III & +0.13$\pm$0.05 & 8.38 & 8 & 9 & h;i \\
         NGC 6633 & 770 & II & -0.08$\pm$0.12 & 8.00 & 10;11 & - & -\\
         Praesepe (NGC 2632) & 750 & II & +0.16$\pm$0.08 & 8.48 & 8 & 12 & b;j;k;l\\
         NGC 6811 & 950 & I &  +0.03$\pm$0.01 & 8.20 & - & - & b;m \\
         NGC 6819 & 2000 & II & +0.09$\pm$0.01 & 8.03 & 13 & - & n \\
         NGC 2420 & 2500 & V & -0.05$\pm$0.02 & 10.68 & 14 & - & - \\
         M 67 (NGC 2682) & 3640 & II & -0.01$\pm$0.04 & 8.96 & 15 & - & -\\
         NGC 2243 & 4000 & VI & -0.38$\pm$0.04 & 10.58 & 16 & - & -\\
         \hline
    \end{tabular}
    \tablefoot{\\ The [Fe/H] values are from \citet{2016A&A...585A.150N} and \citet{2020A&A...643A..71G}, except for UMa \citep{2003ApJ...583..955B} and M 50 \citep{2016ApJ...822...47D}. Distances to the Galactic centre are from \citet{2020A&A...640A...1C}, except for UMa \citep{2020AJ....159..166U} and they are based on the Gaia DR2 measurements. \\ I: \citet{2021arXiv210101183G}; II: \citet{2019A&A...623A.108B}, III: \citet{2016A&A...585A.150N}; IV: \citet{2003ApJ...583..955B}; V: \citet{2020A&A...643A.164S}; VI: \citet{2020A&A...643A..71G}. 1: \citet{2001A&A...372..862R}; 2:\citet{2018A&A...613A..63B}; 3: \citet{2003ApJ...582..410B}; 4: \citet{Balachandran2011} ; 5: \citet{2001ApJ...549..452B}; 6: \citet{2021MNRAS.500.1158J}; 7: \citet{2003ApJ...583..955B}; 8: \citet{2017AJ....153..128C}; 9: \citet{2016ApJ...830...49B}; 10: \citet{1997MNRAS.292..177J}; 11: \citet{2002MNRAS.336.1109J}; 12: \citet{2004ApJ...605..864B}; 13: \citet{2019AJ....158..163D}; 14: \citet{2020A&A...643A.164S}; 15: \citet{2012A&A...541A.150P}; 16: \citet{2013A&A...552A.136F}. a: \citet{2008MNRAS.384..675I}; b: \citet{2021arXiv210101183G}; c: \citet{2009IAUS..258..363I}; d: \citet{2010MNRAS.408..475H}; e: \citet{2009ApJ...695..679M}; f: \citet{2009MNRAS.392.1456I}; g: \citet{2008ApJ...675.1254H}; h: \citet{2011MNRAS.413.2218D}; i: \citet{2016ApJ...822...47D}; j: \citet{2011ApJ...740..110A}; k: \citet{2017ApJ...842...83D}; l: \citet{2019ApJ...879..100D}; m: \citet{2011ApJ...733L...9M}; n: \citet{2015AAS...22544909M}\\
    $\bullet$ [Fe/H] from \citet{2016ApJ...822...47D} \\
    $\star$ [Fe/H] from \citet{2003ApJ...583..955B}\\
    $\ast$ Distance to the Galactic centre from \citet{2020AJ....159..166U}, assuming the Sun is at a distance of 8.34 kpc from the Galactic centre \citep{2020A&A...640A...1C}
}
    \label{tab:OCinfos1}
\end{table*}
  
\subsection{Lithium and Beryllium abundances}
\label{sub:Lisuncluster}
In this work, we consider Li spectroscopic data for a sample of 14 open clusters (see Tab.~\ref{tab:OCinfos1}) with [Fe/H] values between -0.38 and +0.16 dex, and ages between 35 Myrs and 4 Gyrs: IC 2602, IC 2391, Pleiades, $\alpha$ Persei ($\alpha$ Per), M 35, M 50, Coma Berenices (Coma Ber), Ursa Major (UMa), Hyades, NGC 6633, Praesepe, NGC 6819, NGC 2420, M 67, and NGC 2243. We adopted the meteoritic abundance $\rm A(Li) = 3.31$\footnote{$\rm A(X) = log_{10}(N_X/N_H)+12$, where $\rm N_X$ is the number density of element X.} \citep{2009ARA&A..47..481A} as the original abundance of lithium.\\
All the original papers give 1D local thermodynamic equilibrium (LTE) Li abundances, except \citet{2002MNRAS.336.1109J}, where they give non-local thermodynamic equilibrium (NLTE) Li abundances for NGC 6633 stars. As lithium abundances are known to be sensitive to non-LTE effects, we computed the 3D NLTE corrections ($\Delta_{\rm NLTE}$) to all the 1D LTE lithium abundances, using the code \texttt{Breidablik}\footnote{https://github.com/ellawang44/Breidablik} by \citet{2021MNRAS.500.2159W}\footnote{In the case of the NGC 6633 stars from \citet{2002MNRAS.336.1109J}, we first reversed their NLTE correction using the code of~\citet{1994A&A...288..860C} to obtain the 1D LTE abundances and then computed the same 3D NLTE corrections as for the other data.}. To do so, for each star, we used the [Fe/H],  T$_{\rm eff}$, and $\log$g values given in the original papers also providing the lithium abundances (see Tab.~\ref{tab:OCinfos1}). In the specific case of the Pleiades, \citet{2018A&A...613A..63B} do not give the surface gravity. We thus adopted log~$g=4.4$ for its stars, which appears to describe the typical surface gravity of F and G dwarfs\footnote{$\Delta_{\rm NLTE}$ weakly depends on log~$g$ for most of the studied temperature range. For instance, for T$_\mathrm{eff}$=6'000~K, A(Li)=2.5 and [Fe/H]=0, $\Delta_{\rm NLTE}$(log~$g$=4.4)-$\Delta_{\rm NLTE}$(log~$g$=4.0) = -0.0135~dex, which is negligible. However, for T$_\mathrm{eff}\approx$3'900~K to 4'500~K, this difference is more significant and reaches values of about +0.08~dex. In our sample, only 17 out of the 92 stars of Pleiades are in this temperature range, therefore this should not have a significant impact on the result.} well. The absolute values of NLTE corrections rarely exceed 0.1~dex; we consequently do not expect a significant impact on the results. 

We also used the Be spectroscopic data as an additional constraint for the Hyades and Praesepe from \citet{2004ApJ...605..864B} and \citet{2016ApJ...830...49B}. Observations were directly extracted from \citet{2004ApJ...605..864B,2016ApJ...830...49B} without any modification. We adopted the meteoritic abundance $\rm A(^9Be)=1.41$ \citep[][]{2009ARA&A..47..481A} as the original abundance of beryllium.

\subsection{Age}
\label{subsect:age}
There is no self-consistent determination in the literature of the ages of all the clusters we consider here, and this task is out of the scope of this work. To be consistent with Paper I, we used the ages from \citet[][when available]{2019A&A...623A.108B} for the clusters with Li abundance measurements. For the clusters with rotation period measurements, we used the ages from \citet[][when available]{2021arXiv210101183G}. For the five clusters with Li data that were not studied in those two works, we took the ages quoted in the respective observation papers (see Table~\ref{sec:obs}). Except for the Pleiades, the differences in age determinations from the literature weakly affect our conclusions (see discussion in  Sect.~\ref{sect:protmass}).

\subsection{Effective temperature $T_{\rm eff}$}
\label{subsect:teff}
The effective temperatures used in this study were taken from the same original sources as the lithium (or beryllium) abundances for consistency (see Tab.~\ref{tab:OCinfos1}). In rare cases, spectroscopic T$_{\rm eff}$ are available (this is the case of M67 and NGC 2420). In most cases, T$_{\rm eff}$ were determined using colour-temperature calibration relations from (B-V), (V-I$_c$), or (V-K) colour indices. For the 14 open clusters presented in Table~\ref{tab:OCinfos1}, at least six different relations and methods were used. 
These relations differ by their metallicity dependence, the zero-point of their temperature scale, the colours used, etc. The relative consistency between the different calibration relations was tested in several previous studies \citep{2015MNRAS.454.2863H,2010A&A...512A..54C,2010A&A...522A..98M}. These studies emphasise that maximum differences of about 100 K can be found when applying the different temperature scales to dwarf stars in the metallicity and photometric domain considered here. We thus consider that the general agreement between the different temperature scales adopted in the original papers that we use in this work is satisfactory.

\subsection{Surface and internal rotation} 
\label{subsub:rotationprofile}

 We compared the surface rotation predicted by our models with the observational data set gathered by \citet{2009IAUS..258..363I}, \citet{2015A&A...577A..98G}, and \citet{2021arXiv210101183G} for low-mass stars among a sample of 12 open clusters with ages between 5 Myrs and 2 Gyrs: NGC 2362, NGC 2547, IC 2391, $\alpha$Per, Pleiades, M 35, M 50, M 37, Hyades, Praesepe, NGC 6811, and NGC 6819 (see details in Table~\ref{tab:OCinfos1}).
 
We constrained our model predictions for internal rotation using the asteroseismic measurements obtained by \citet[][]{2015MNRAS.452.2654B} for a sample of MS field stars observed by \textit{Kepler}. We selected a sub-sample of four stars with metallicities close to that of the Hyades.
 
\section{Stellar evolution models -- Input micro- and macro-physics}
\label{STAREVOL}

\subsection{General assumptions}
\label{subsection:Generalities}

We followed the conclusions of Paper I on different transport processes of chemicals and angular momentum (namely: rotation, penetrative convection, parametric turbulence, parametric viscosity, and atomic diffusion, as described below), which were tested with respect to Li abundances and surface and internal rotation constraints for solar-type stars. 
We used the nomenclature described in Paper I. For instance, model $^M_{\nu}R1^{T6.42}_A$ is a model computed with an R1 prescription for rotation-induced turbulence and parametric viscosity ($\nu$=$\nu_{\rm add}$), penetrative convection (A), parametric turbulence down to $\log T_0$ = 6.42, at median initial rotation velocity (M), and with atomic diffusion. All the details about the corresponding prescriptions for the different processes are given below.\\  
 To a broader range of masses and metallicities, we  applied the prescriptions that were identified to be the most relevant for solar-type stars. Models were computed in the mass range between 0.8 M$_\odot$ and 1.5 M$_\odot$ ($\delta \mathrm{M} = 0.1 \mathrm{M}_\odot$) for seven values of [Fe/H] (-0.4, -0.2, -0.1, -0.05, 0, +0.10, +0.15) that cover the metallicity range of the Galactic open clusters described in Table~\ref{tab:OCinfos1}. Computations started on the Hayashi track at the beginning of the deuterium burning phase 
on the PMS that we consider as the time zero of the evolution. 

\subsection{Input physics}
\label{sub:inputphysics}
We used the same updated version of the stellar evolution code STAREVOL as in Paper I
\citep[for general information and previous versions see][]{2000A&A...358..593S,2006A&A...453..261P,2009A&A...495..271D,2012A&A...543A.108L,2019A&A...631A..77A}, which we refer to for details and references. 
The models presented in this work were computed with the same inputs physics (equation of state, opacities, nuclear reactions, model atmosphere, and mass loss).
We used the same values for the mixing length parameter ($\alpha_{\rm MLT}$, assuming the Schwarzschild criteria for convective stability) and the initial abundances that resulted from model calibrations on the Sun that were carried out for classical models (no transport of chemicals beyond convection), and for models including both rotation and atomic diffusion (respectively models C and R1 from Paper I).

\subsection{Atomic diffusion, penetrative convection, parametric turbulence}
\label{sub:dmicdadturb}
Atomic diffusion was implemented according to \citet{1986ApJS...61..177P} and \citet{1994ApJ...421..828T}. We did not take radiative accelerations into account. Their impact starts, however, to be non-negligible for stars with effective temperature higher than $\sim$ 6'800~K, which corresponds to $\sim$ 1.4~M$_{\odot}$ at solar metallicity \citep[][]{1998ApJ...492..833R,2002ApJ...568..979R,2018A&A...618A..10D,2020A&A...633A..23D}. This is discussed in Sect.~\ref{sect:CONCLUSION}.\\

Penetrative convection was treated as an overshoot and computed following the formalism of \citet{2019ApJ...874...83A} with the following expression for the diffusion coefficient:
\begin{equation}
    D_{A}(r) \approx D_0 \left[1-\exp\left(-\exp\left(\frac{r-r_{bcz}}{d_{ov} \times \left(\rm{\frac{v}{v_0}}\right)^{3/2}} + \frac{\mu}{\lambda} \right)\right)\right], 
    \label{eq:dkyle}
\end{equation}
where $D_0 = (\upsilon_{\rm conv} \times H_p \times \alpha_{\rm MLT})/3$ is the convective turbulent diffusivity, with $\upsilon_{\rm conv}$ being the mean velocity of the convective elements obtained from MLT, $\alpha_{\rm MLT}$ being the mixing length parameter, and $H_p$ being the pressure scale height; $r$ is the local radius; $r_{bcz}$ is the radius at the base of the convective zone; $\rm{(v/v_0)}$ is the ratio of the velocity of the convective elements when taking rotation into account for the non-rotating inviscid value; and $d_{ov}=0.0325$ (as calibrated by Paper I to reproduce the Li abundance in solar-type stars in very young open clusters). The coefficients $\lambda = 6\times10^{-3}$ and $\mu = 5\times10^{-3}$ are as prescribed by \citet{2017ApJ...845L...6B} based on the simulations of \citet{2017A&A...604A.125P}.\\ 

Parametric turbulence is defined according to \citet{2000ApJ...529..338R} and \citet{2005ApJ...619..538R} with the following prescription for the diffusion coefficient:
\begin{equation}
    D_{\rm{T_0}} = 400 D_{\rm{He}}(T_0)\left[\frac{\rho(T_0)}{\rho}\right]^3,
\label{eq:dturb1}
\end{equation}
where $T_0$ is a free parameter that sets the depth of the maximum efficiency of the mixing depending on the value of the atomic diffusion coefficient He ($D_{\rm{He}}$). Initially introduced by \cite{2000ApJ...529..338R} to counteract an impact of atomic diffusion in AmFm stars that was too strong, we introduced  it in our models with rotation to increase the surface Li depletion predicted on the MS and reproduce the observations for solar-type stars and for the Sun. For all the reference models, we adopted $\log T_0 = 6.42$ as calibrated by Paper I. 
We also present a set of models for the Hyades metallicity ([Fe/H] = + 0.15 dex), where we increased this depth to $\log T_0$ = 6.5. 

\begin{figure*}[h!]
         \center
         \includegraphics [width=150mm]{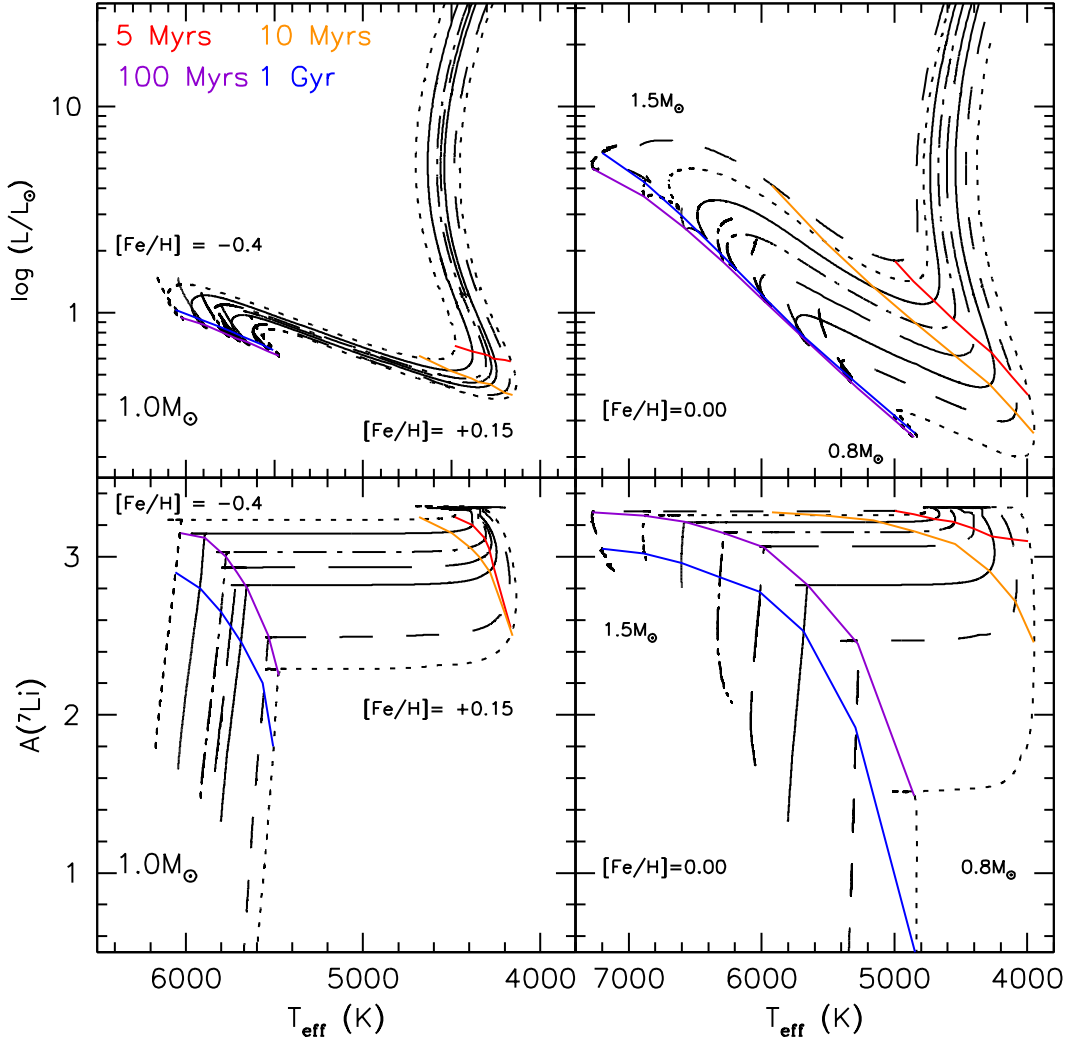}
         \caption{(Top) Evolutionary tracks in the Hertzsprung-Russell diagram of $^M_{\nu}R1^{T6.42}_A$ models of 1.0$M_{\odot}$ computed for seven values of [Fe/H] (Left) and of different masses for [Fe/H]=0 (Right). (Bottom) Surface Li abundance for the corresponding models. The coloured-lines connect points of the same age.
         }
         \label{fig:HrdLimulti}
 \end{figure*}

\subsection{Angular momentum evolution and rotation-induced mixing}
\label{sub:stellarrotation}
Stellar rotation was implemented in STAREVOL as described by \citet{2016A&A...587A.105A,2019A&A...631A..77A} and Paper I. We adopted the shellular rotation hypothesis developed by \citet{1992A&A...265..115Z}, \citet{1998A&A...334.1000M}, and \citet{2004A&A...425..229M} to describe the transport of angular momentum and chemicals by meridional circulation, treated as an advective process for angular momentum, and turbulent shear (vertical and horizontal), treated as diffusive processes. We followed \citet{2012A&A...544L...4E,2019A&A...621A..66E} and \citet{2016A&A...589A..23S} who introduced an additional parametric vertical viscosity $\nu_{\rm add}$ in order to flatten the internal rotation profile as evidenced by helio- and asterosismology of low-mass stars. 
The transport of angular momentum hence obeys the following advection-diffusion equation:
\begin{equation}
    \rho \frac{d}{dt}(r^2 \Omega) = \frac{1}{5 r^2} \frac{\partial}{\partial r} (\rho r^4 \Omega U_2) + \frac{1}{r^2} \frac{\partial}{\partial r}\left((\nu_v+\nu_{\rm{add}}) r^4 \frac{\partial \Omega}{\partial r}\right),
    \label{eqrot}
\end{equation}

where $\rho$, $r$, $\Omega$, $U_2$, and $\nu_v$ are the density, the radius, the angular velocity, the meridional circulation velocity, and the vertical shellular component of the turbulent viscosity, respectively. For the reference models, we adopted $\nu_{\rm{add}}=3.5\times10^4 \rm{cm^2s^{-1}}$ in the entire radiative region. This value was calibrated in Paper I on the solar angular velocity profile provided by helioseismology. We also present a set of models for the Hyades metallicity ([Fe/H] = + 0.15 dex) where we modified this value depending on the initial stellar mass ($\nu'_{\rm{add}}=3.5\times10^5$, $1.0\times10^5$, $2.5\times10^5$, $6.5\times10^5$, and $8.5\times10^5$ $\rm{cm^2s^{-1}}$ for the 1.5, 1.4, 1.35, 1.3, and 1.2 M$_{\odot}$ models, respectively). These values were kept constant during the entire evolution.\\

We explored two combinations of prescriptions for turbulence shear referred to as R1 and R2 as in Paper I; R1 includes prescriptions from \citet{2018A&A...620A..22M} and \citet{1992A&A...265..115Z} for the horizontal diffusivity $D_h$ and the vertical diffusivity $D_v$, respectively, and R2 includes prescriptions from \citet{1992A&A...265..115Z} and \citet{1997A&A...317..749T} for $D_h$ and $D_v$, respectively. The detailed expressions of the four turbulent diffusion coefficients can be found in Appendix~B of Paper I. 
We used the same parameters as in Paper I for the extraction of angular momentum at the stellar surface due to magnetised winds (m=0.22, p=2.1, $\chi=14$, $K=7.5\times10^{30}$ erg unless otherwise indicated) according to the \citet{Matt2015} formalism. Models were computed for three values of the initial rotation period on the PMS: 1.6, 4.5, and 9.0 days, which are referred to as fast ($^{\textbf{F}}R$), median ($^{\textbf{M}}R$), and slow ($^{\textbf{S}}R$) rotating models, respectively. The disc coupling timescale was set at $\tau_{\rm disc}$ = 2.5 Myrs for the fast rotators and at $\tau_{\rm disc}$ = 5 Myrs for the median and the slow rotators, in agreement with \citet{2015A&A...577A..98G} and \cite{2019A&A...631A..77A}.

To summarise, the best models for solar-type stars that we developed in Paper I include the self-consistent treatment of physical processes as well as parametrised additional transports for angular momentum and chemicals as summarised in Tab.~\ref{tab:Modelbest}. The parameters are strongly constrained by observations (see col. 3 Tab.~\ref{tab:Modelbest}) and cannot be varied freely. In the present study, we explore a possible variation in the efficiency of these additional processes with metallicity and initial mass.
\begin{table*}[t!]
    \centering
    \caption{Transport processes considered in this work.}
    \begin{tabular}{c|c|c|c}
         Process & Quantity & Adjusted parameter & Observational constraints \\
         \hline \hline
         Atomic diffusion & C & --  & -- \\
         Meridional circulation & C \& AM & -- & -- \\
         Shear-induced turbulence (R1 or R2) & C \& AM & -- & -- \\
         Magnetic torque & AM & $K = 7.5\times10^{30}$ erg & 
         Surface rotation of solar-type stars \\
         Penetrative convection & C & $d_{ov} = 0.0325$ & 
         Surface Li abundance of young solar-type stars\\
         Parametric turbulence & C & log T = 6.42 & 
         Surface Li abundance of MS solar-type stars \\
         Vertical viscosity & AM & $\nu_{\rm add} = 3.5\times10^{4}$ $\rm cm^2s^{-1}$ & Solar rotation profile \\
         \hline
    \end{tabular}
    \label{tab:Modelbest}
    \tablefoot{\\ Transport processes (Column 1) and values of the free parameters (when relevant) (Column 3) with the corresponding observational constraints adjusted (Column 4) for the best model for solar-type stars (Paper~I) and adopted in this study. The flags in Column 2 indicate the transported quantity (C for chemicals and AM for angular momentum).}
\end{table*}

\section{Model predictions for Li, Be, and surface rotation - The Hyades and Praesepe test case}
\label{sect:predictions}
 
In this section, we explore the predictions of model $_{\nu}R1^{T6.42-6.5}_A$ over a range of masses and metallicities for three different rotation rates. We recall that this model was developed for the specific case of solar-type stars in Paper I. 
We discuss the general impact of mass and metallicity on Li depletion, and then we compare the predictions for the behaviour of surface Li, Be, and rotation rates to the observations in the Hyades and Praesepe, and to the predictions of classical models and of model $^M_{\nu}R2^{T6.42}_A$ computed with different prescriptions for rotation-induced turbulence (see Tab.~\ref{tab:Modelinfos2}). 

\begin{figure}[t]
         \center
         \includegraphics [width=90mm]{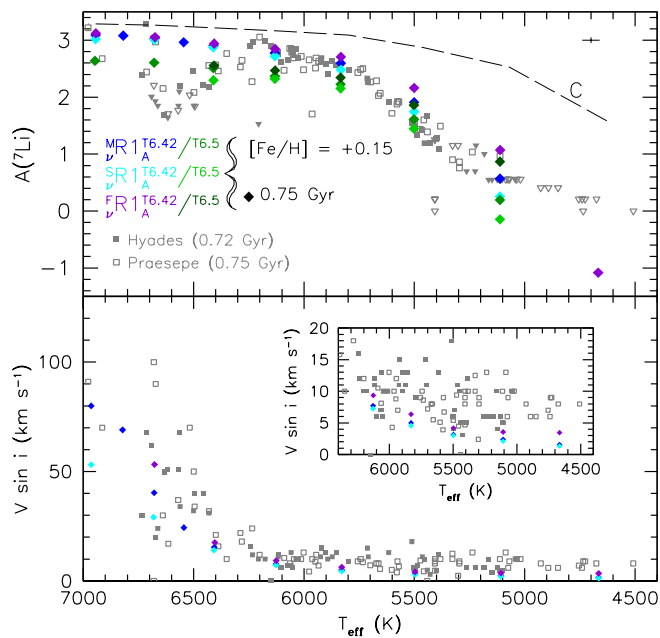}
         \caption{\textit{Top}: Li surface abundance versus T$_\mathrm{eff}$ in the Hyades (filled squares) and Praesepe (open squares) from \citet{2017AJ....153..128C}; downward triangles are for upper limits. Comparison to the predictions at 0.75 Gyr (diamonds) of models with [Fe/H] = +0.15 computed under several assumptions: classical models ($C$; dashed black line) and complete models $_{\nu}R1^{T6.42}_A$ (cyan, blue, and violet diamonds for slow, median, and fast initial rotation velocities, respectively) and $_{\nu}R1^{T6.5}_A$ (light green, green, and dark green diamonds for slow, median, and fast initial rotation velocities, respectively). \textit{Bottom}: V~sin~i versus T$_\mathrm{eff}$ in the Hyades (filled squares) and Praesepe (open squares) from \citet{2017AJ....153..128C}. Comparison to the surface rotation velocity predictions of model $_{\nu}R1^{T6.42}_A$ at 0.75 Gyr. Insert: Zoom on the cool side of the Kraft rotation break including model predictions for 0.8-1.2$M_{\odot}$ stars.
         }
         \label{fig:LiTeffFeH015}
 \end{figure}
 
\subsection{Impact of metallicity and mass on Li evolution}
\label{subsect:general}

Figure~\ref{fig:HrdLimulti} shows the evolutionary tracks in the Hertzsprung-Russell diagram (PMS and MS) and the predicted surface Li abundance as a function of T$_{\rm eff}$ for a selection of models (all computed with median rotation). 
We see the well-known impact of varying the mass and the metallicity on the evolution tracks, and the global consequence on Li depletion. When metallicity decreases for a given mass, or when the mass increases for a given metallicity, the model is hotter and brighter. As a consequence, its convective envelope retracts more rapidly at the beginning of the Hayashi track on the PMS, and it is thinner on the MS. Hence, its base is more distant from the layers where Li is burning, which overall leads to lower Li depletion, although the quantitative details depend on the efficiency of the transport process(es) that connect(s) the base of the convective envelope to the Li-free region (see below).

In all the models shown in Fig.~\ref{fig:HrdLimulti}, Li decreases along two successive episodes, first on the PMS, and later along the MS. The PMS depletion episode is first due to Li burning in the convective envelope at the beginning of the Hayashi track and later to overshoots that are both dependent on the size of the convective envelope and its timescale for retraction along the PMS. As a consequence, PMS depletion is minute for stars with ZAMS\footnote{Zero age main sequence.} T$_{\rm eff}$ higher than $\sim$ 6'500 K, and it increases with decreasing stellar mass and increasing metallicity.\\ 
The MS depletion episode is due to the combined effects of rotation-induced mixing and parametric turbulence. The first process is more efficient in cooler stars (i.e. less massive at a given metallicity, or more metal-rich at a given mass), which undergo more significant extraction of angular momentum by their magnetised winds (see e.g. Fig.~6 in \citealt{2019A&A...631A..77A}). Similar mass and metallicity dependencies exist in the case of the second process because in all of our models we assume, due to Eq.~(\ref{eq:dturb1}), the same value for the free parameter $T_0$ which is closer to the temperature of the base of the convective envelope in less massive or more metal-rich stars. In summary, overall Li depletion is stronger in less massive, more metal-rich stars.

\subsection{The Hyades and Praesepe}
\label{subsect:hyades}
\begin{table*}[t!]
    \centering
    \caption{Models computed for the specific case of the Hyades and Praesepe open clusters at [Fe/H] = +0.15.}
    \begin{tabular}{c|c|c|c}
         Model & $\nu_{\rm add}$ $\rm (cm^2.s^{-1})$ & K ($10^{30}$ erg) & Turbulence\\
         \hline \hline
         $C$ & 0 & $7.5$ & none \\
         $_{\nu}R1^{T6.42}_A$ & $3.5\times10^4$ & $7.5$ & $D_{T6.42}$\\
         $_{\nu}R1^{T6.5}_A$ & $3.5\times10^4$ & $7.5$ & $D_{T6.5}$\\
         $^M_{\nu}R2^{T6.42}_A$ & $3.5\times10^4$ & $7.5$ & $D_{T6.42}$\\
         $^M_{\nu'}R2^{T6.42}_A$ & \footnotesize{$1.0- 2.5 - 3.5 - 6.5 - 8.5\times10^5$} & $7.5$ & $D_{T6.42}$\\
          $^M_{\nu}R2^{T6.42}_{A.K'}$ & $3.5\times10^4$ & $1.5$ & $D_{T6.42}$\\
         \hline
    \end{tabular}
    \label{tab:Modelinfos2}
\end{table*}
Here we focus on the well-studied Hyades and Praesepe for which Li, Be, and surface rotation data are available. These two open clusters are close in age (0.72 Gyr and 0.75~Gyr, respectively) and metallicity ([Fe/H] = +0.13$\pm$0.05 and +0.16$\pm$0.08~dex, respectively) according to the references quoted in Table~\ref{tab:OCinfos1}.

Figure~\ref{fig:LiTeffFeH015} shows the observed Li abundances and rotation velocities (V~sin i) as a function of the effective temperature from \citet{2017AJ....153..128C}. Both clusters exhibit similar patterns, namely the so-called Li dip between $\sim$ 6'400 and 6'800 K, the Li-T$_{\rm eff}$ decrease in G-type stars on the cool side of the dip, and the well-know break in rotation velocity ($\approx$ 6'400-6'500~K) for dwarf stars later than the F4 spectral type  \citep{Schatzman1959,Schatzman1962,1967ApJ...150..551K,1987PASP...99.1067B}. In the same figure, we show the predictions at 0.75~Gyr of the classical models, and of the complete models $_{\nu}R1^{T6.42}_A$ and of $_{\nu}R1^{T6.5}_A$ models computed with [Fe/H] = +0.15 for three different initial rotation velocities (slow, median, and fast) for masses between 0.8 and 1.5 $M_{\odot}$ (mass step 0.1 $M_{\odot}$), and for a median initial rotation velocity for the 
1.35 and 1.45~$M_{\odot}$ models. 

 \begin{figure*}[h!]
         \center
         \includegraphics [width=150mm]{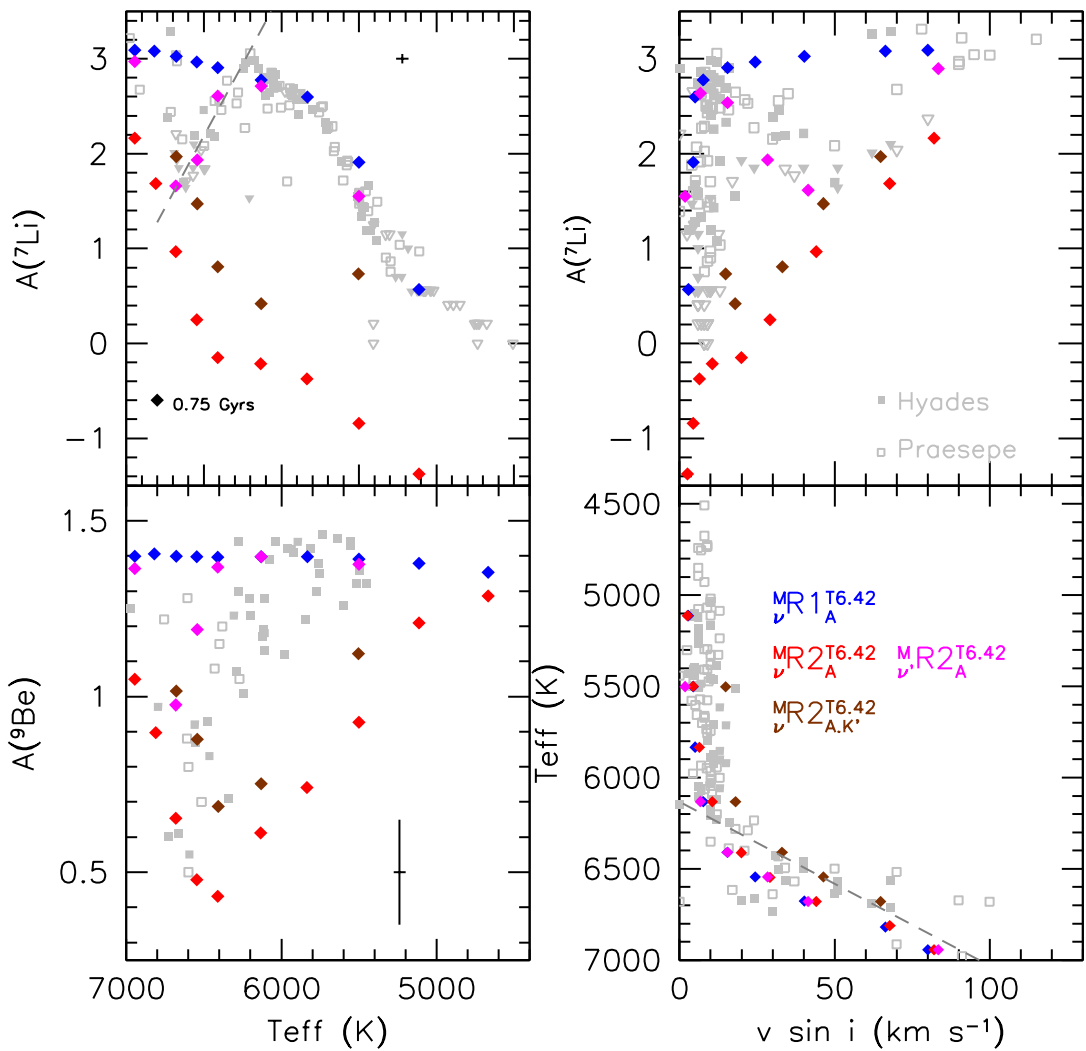}
         \caption{Comparison of the Li, Be, and V~sin i observations in the Hyades and Praesepe (filled and open grey squares, respectively; triangles are for abundance upper limits; see Table~\ref{tab:OCinfos1} for references) with the predictions of models $^M_{\nu}R1^{T6.42}_A$ (blue diamonds) and $^M_{\nu'}R2^{T6.42}_A$ for different values of $\nu'_{\rm add}$ and K (colour-coded, see also Tab.~\ref{tab:Modelinfos2}). Model predictions are shown at 0.75~Gyr. The grey dashed line represents the empirical relation in the cold edge of the dip obtained by \citet{2017AJ....153..128C}.
         }
         \label{fig:LiTeffFeH015bis}
 \end{figure*}

As has been long known in the literature, and as shown in Fig.~\ref{fig:LiTeffFeH015}, classical models (model C) that account solely for convection predict Li depletion on the PMS only, when the size of the convective envelope is large enough to reach the Li-burning temperatures along the Hayashi track (see Sect.~\ref{subsect:general}). While the predicted mass and T$_{\rm eff}$ dependence of the Li abundance after the PMS depletion is similar to the observed pattern for G-type stars, the predicted surface Li abundances are too high compared to the observations in the Hyades and Praesepe, as well as in the other clusters of different ages considered in this work (see Sect.~\ref{sect:multicluster}).
 
Models $_{\nu}R1^{T6.42}_A$  and $_{\nu}R1^{T6.5}_A$ predict rotation velocities that account for the observed V~sin~i trends well (including the position in effective temperature of the Kraft break) in both open clusters, thanks to the extraction of angular momentum due to magnetised winds (see also Fig.~\ref{fig:Protmass}). However, on the cool side of the Kraft rotation break (see insert in Fig.~\ref{fig:LiTeffFeH015}), models reproduce the lower observational envelope well, but not the entire observed spread. It is mainly the case for the lowest stellar masses where the parametric viscosity is more efficient. 

\begin{figure}[h!]
         \center
         \includegraphics [width=80mm]{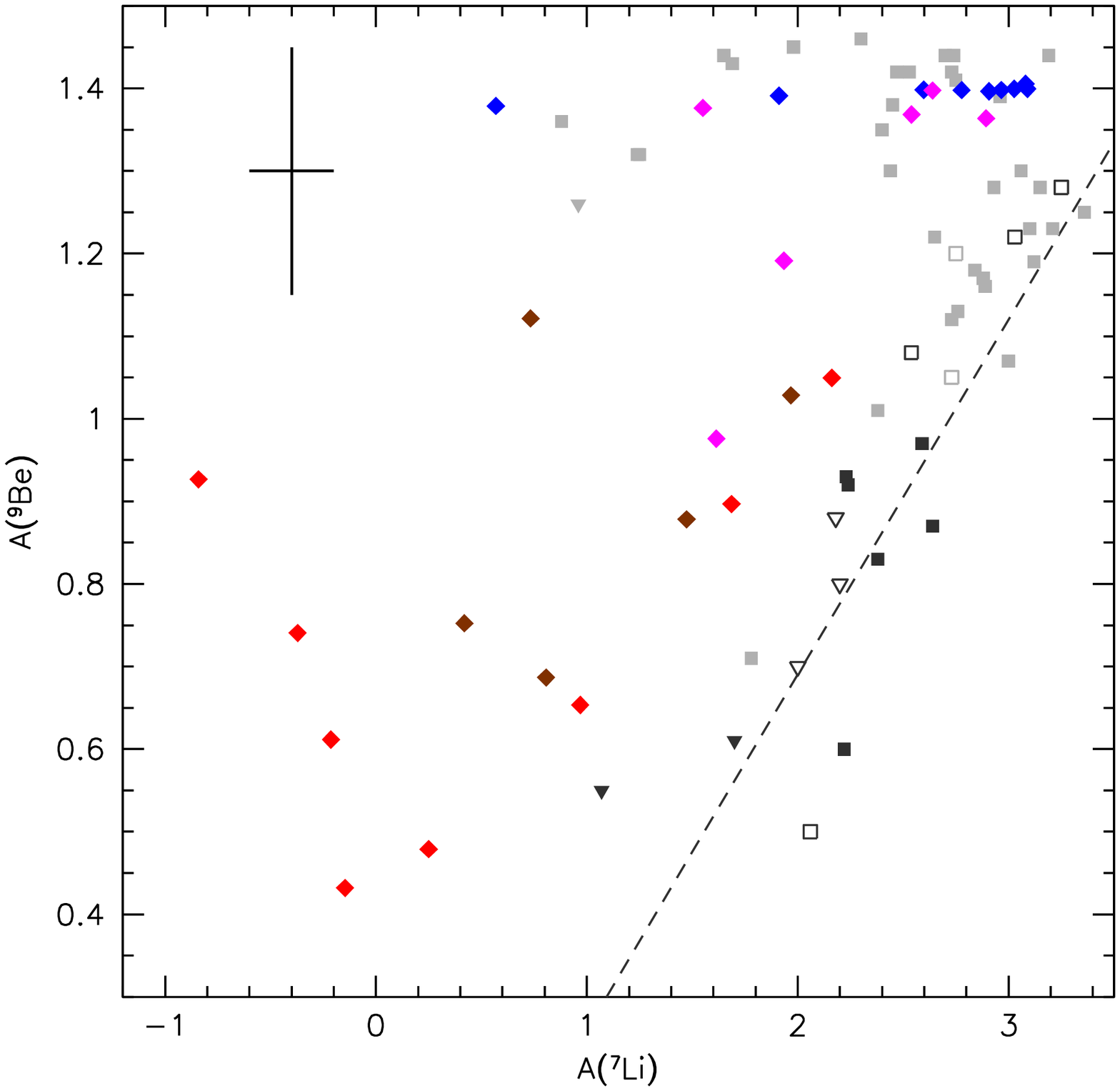}
         \caption{Comparison of the models $^M_{\nu}R1^{T6.42}_A$ (blue diamonds), $^M_{\nu}R2^{T6.42}_A$ (red diamonds), $^M_{\nu}R2^{T6.42}_{A.K'}$ (brown diamonds), and $^M_{\nu'}R2^{T6.42}_A$ (magenta diamonds) with observations of surface Li and Be in the Hyades (filled squares) and Praesepe (open squares) open clusters. Downward triangles are for Li upper limits. Stars corresponding to the Li dip (6'400 K < $T_{\rm eff}$ < 6'800) are represented in dark grey and stars outside of the Li dip are represented in light grey. Observations are directly extracted from \citet{2004ApJ...605..864B,2016ApJ...830...49B} without any modification (no NLTE correction for Li abundances). The dark grey dashed line represents the linear correlations found by \citet{2020ApJ...888...28B} with $\rm A(Be) = 0.43\times A(Li) - 0.17$ for stars in the cold edge of the Li dip.}
         \label{fig:LivsBe}
 \end{figure}

\begin{figure*}[h!]
         \center
         \includegraphics [width=150mm]{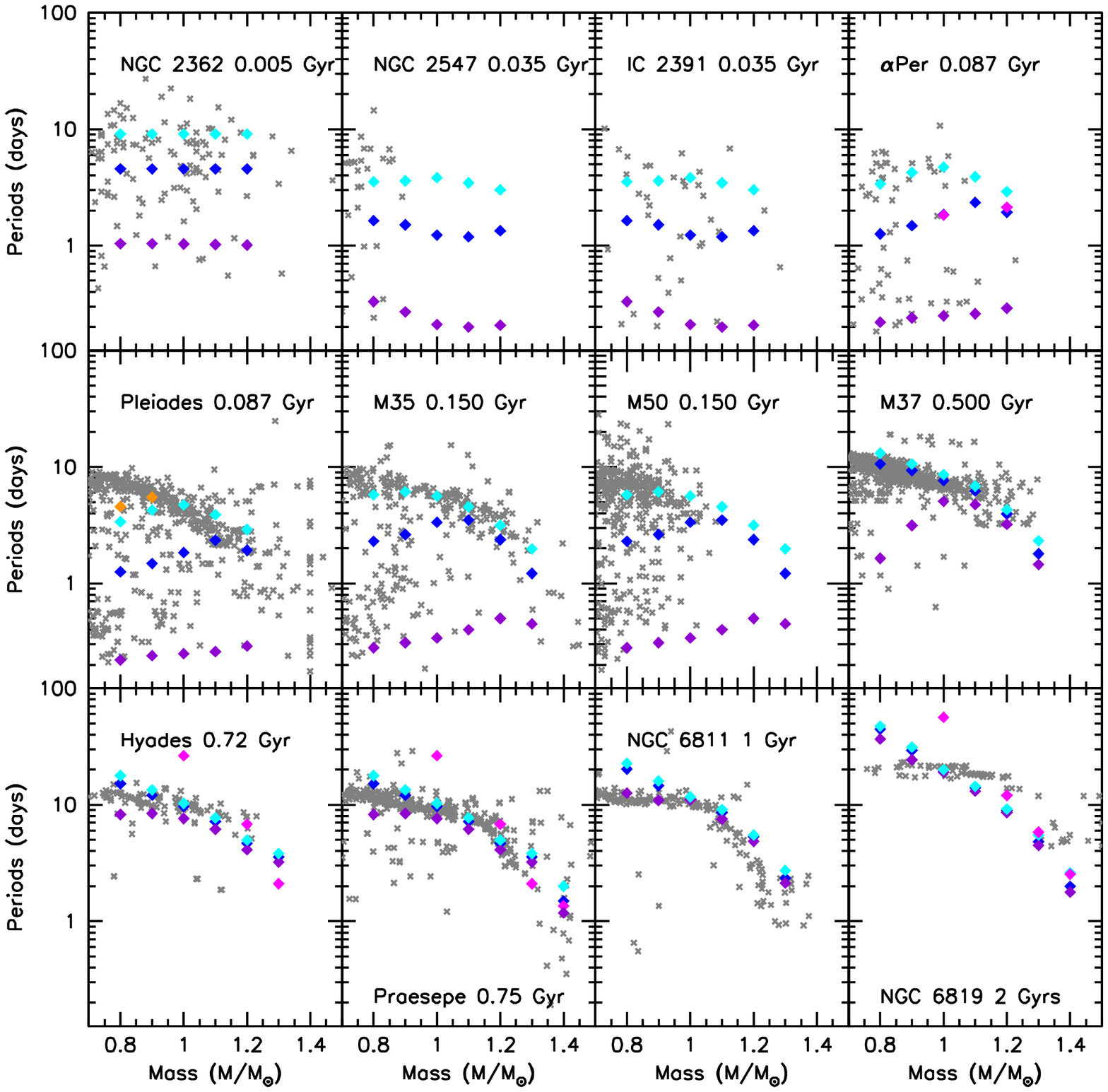}
         \caption{Surface rotational period versus mass. Comparison of the models $_{\nu}R1^{T6.42}_A$ (diamonds) at three different initial velocities: slow (cyan), median (blue), and fast (violet) with the observations for open clusters at different ages and metallicities. The orange diamond for Pleiades corresponds to a slow rotating model at 125 Myrs. In the case of $\alpha$Per, Hyades, Praesepe, and NGC 6819, we also show the predictions of the $_{\nu'}^{M}R2^{T6.42}_A$ models (magenta diamonds). Observations' references are reported in Table~\ref{tab:OCinfos1}. }
         \label{fig:Protmass}
 \end{figure*}

Models $_{\nu}R1^{T6.42}_A$ and $_{\nu}R1^{T6.5}_A$ also provide a very good fit to the Li data in G-type stars on the cool side of the Li dip where angular momentum extraction is maximal, that is, for stellar masses lower than 1.3$M_{\odot}$ for the metallicity of these clusters. At the age of the Hyades and Praesepe, the increased (with respect to the classical model C) Li depletion is mainly due to penetrative convection (Eq.~\ref{eq:dkyle}) along the PMS, and only slightly due to turbulent mixing at the beginning of the MS. As discussed in Paper I, the \citet{2019ApJ...874...83A} expression for penetrative convection leads to larger Li depletion for slow rotators than for fast rotators because of the influence of the rotation rate on the depth of the overshoot via the ratio $v/v_0$ in Eq.~(\ref{eq:dkyle}). This anti-correlation with the rotation rate, which has long been observed \citep[e.g.][]{2008A&A...489L..53B,2018A&A...613A..63B,2020A&A...635L..13A}, is obtained here for the first time for the entire mass range considered at the Hyades and Praesepe age. \footnote{This anti-correlation was initially obtained by \citet{2015MNRAS.449.4131S} who invoked a radius inflation dependent on the rotation velocity to explain the anti-correlation at the Pleiades age. However, even though it has been confirmed in the Pleiades \citep[][]{2016ApJ...829...32S,2017AJ....153..101S}, the effect of this process still needs to be clarified at older ages than Pleiades and for different metallicities \citep[e.g.][]{2019ApJ...879...39J,2019MNRAS.483.1125J}.} In addition, the predicted Li spread induced by the different initial rotation rates assumed here (Sect.~\ref{sub:stellarrotation}) is amplified in lower mass stars where the base of the convective envelope is deeper and closer to the Li burning layers, which also leads to more efficient parametric turbulence on the lower mass end. For this later process, we present model predictions for two values of the parameter $T_0$ from Eq.~(\ref{eq:dturb1}), that is, $\log T_0$ = 6.42 as calibrated in the Sun and $\log T_0$ = 6.5 which better fits the data spread in cool stars of the Hyades and Praesepe.  
Overall, our complete model $_{\nu}R1^{T6.42-6.5}_A$ reproduces fairly well the Li-$T_{\rm eff}$-V ${\sin i}$ trend and the observed Li dispersion on the cool side of the Li dip (i.e., G-type and late F-type stars). In this T$_{\rm eff}$ domain, it also accounts for the Be data in the Hyades and Praesepe, as shown in Figs.~\ref{fig:LiTeffFeH015bis} and ~\ref{fig:LivsBe}. 

However, models $_{\nu}R1^{T6.42-6.5}_A$ predict little Li depletion (both on the PMS and the MS at the ages of the Hyades and Praesepe) for stars more massive than 1.3$M_{\odot}$ that have small convection zones that are vanishing on the MS and which undergo a negligible angular momentum braking. As a consequence, it does not reproduce the depth of the Li dip, nor the Li rise on the cool edge of the dip. Around 6'600 K, where many stars only have Li upper limits, the maximum Li depletion predicted reaches only $\sim$ 1~dex in the case where we assume slightly deeper parametric turbulence (i.e. $\log T_0 = 6.5$).  
This low depletion mainly results from the choice of prescriptions for the horizontal and vertical diffusivities $D_h$ and $D_v$ (taken here from \citealt{2018A&A...620A..22M} and \citealt{1992A&A...265..115Z}, respectively; see Sect.~\ref{sub:stellarrotation}). This is in contrast with previous studies on the hot edge of the Li dip \citep{2003A&A...399..603P}, where the use of the prescriptions by \citet{1992A&A...265..115Z} and \citet{1997A&A...317..749T} for D$_h$ and D$_v$, respectively, led to a good match between model predictions and observations. As can be seen in Fig.~\ref{fig:LiTeffFeH015bis}, much stronger Li and Be depletion is indeed obtained in the model $^M_{\nu}R2^{T6.42}_A$ where we used the same prescriptions as in that earlier paper. 
As already evidenced in Paper I for the solar case, this combination however leads to Li depletion that is too strong in F- and G-type stars for the value of $\nu_{\rm add}$ adopted in that paper and here ($3.5\times10^4$ ${\rm cm^2.s^{-1}}$).
Additionally, reducing the torque parameter K for the wind from \citet[][]{Matt2015}, from $7.5\times10^{30}$erg to $1.5\times10^{30}$erg, improves the comparison on the hot edge of the Li dip and of the Be dip, but not on the cool edge as can be seen in Figs.~\ref{fig:LiTeffFeH015bis} and  ~\ref{fig:LivsBe} for model $^M_{\nu}R2^{T6.42}_{A.K'}$.

\citet{1998ASPC..154..973T} reached similar conclusions as \citet{2003A&A...399..603P}, using the same turbulence prescriptions ($D_h$ and $D_v$ from \citealt{1992A&A...265..115Z} and \citealt{1997A&A...317..749T}, respectively). These authors interpreted the rise in Li on the cool edge of the dip as the signature of the appearance of a process that transports angular momentum more efficiently than meridional circulation and turbulent shear. 
\citet{2003A&A...405.1025T,2005A&A...440..981T} then showed that the generation of internal gravity waves (hereafter IGW) by the stellar convective envelope becomes efficient inside the Li dip and increases on its cool edge. In their model, the maximum efficiency of the IGWs in terms of angular momentum transport is expected at T$_{\rm eff}$ around 5'800-5'900~K, before it decreases in cooler stars (see Fig.~6 of \citealt{2003A&A...405.1025T} and Fig.~4 of \citealt{2004A&A...418.1051T}). Last but not least, their model also accounts for the internal solar rotation \citep{2005Sci...309.2189C}. It thus has the proper mass and T$_{\rm eff}$ dependence to be the required transport mechanism involved, contrary to other processes that could also potentially transport angular momentum \citep[e.g. the Taylor-Spruit dynamo][]{2002A&A...381..923S,2005A&A...440L...9E,2010A&A...519A.116E}. 

Following this theoretical trend for IGWs, we computed the $^M_{\nu'}R2^{T6.42}_A$ of various masses assuming a T$_{\rm eff}$ dependence of the parametric viscosity $\nu'_{\rm add}$ (increasing value with decreasing mass on the cool edge of the Li dip) in order to mimic a transport of angular momentum as predicted for low-mass stars.
Although the use of a uniform and constant parametric viscosity within the radiative layers leads to a very different angular momentum profile than that shaped by IGW (compare Fig.\ref{fig:profomega} with Fig.1 of \citealt{2005Sci...309.2189C}), higher values of $\nu'_{\rm add}$ mimic a more efficient internal transport of angular momentum compared to that driven by meridional circulation and turbulent shear (see 
Fig.~\ref{fig:profomega} which is discussed in detail in Sect.~\ref{sect:CONCLUSION}), hence reducing their efficiency for the transport of chemicals.
The $\nu'_{\rm add}$ values required to reproduce the shape of the Li dip are
$3.5\times10^{5}$ ${\rm cm^2.s^{-1}}$ and $1.0\times10^{5}$ ${\rm cm^2.s^{-1}}$ for the 1.5 and 1.4~$M_{\odot}$ models, respectively, on the hot edge of the Li dip, and $2.5\times10^{5}$, $6.5\times10^{5}$, and $8.5\times10^{5}$ ${\rm cm^2.s^{-1}}$ for the 1.35~$M_{\odot}$, 1.3~$M_{\odot}$, and 1.2~$M_{\odot}$ models, respectively, on the cool edge of the dip.
On the other hand, the Li abundance in the 1.0~$M_{\odot}$ model is well reproduced for $\nu'_{\rm add} = 2.5\times10^{5} {\rm cm^2.s^{-1}}$. However for this mass, it leads to a reduction that is  too strong of the surface velocity (see Figs.~\ref{fig:Protmass} and \ref{fig:profomega}), which makes it hard to reconcile with $P_{\rm rot}$ measurements in the Hyades and Praesepe. 
As shown in Figs.~\ref{fig:LiTeffFeH015bis} and \ref{fig:LivsBe}, these assumptions explain the Be plateau in cool stars, within the observational uncertainties. However, the $^M_{\nu'}R2^{T6.42}_A$ models do not account for the large Be depletion in the Be dip nor in its cool edge. This conclusion holds for all the initial rotation rates explored here. 

It thus appears very difficult to reconcile the Li, Be, and surface rotation rate constraints in the test case of the Hyades and Praesepe. While the $_{\nu}R1_A^{T6.42}$ models account for the surface rotation well over the entire mass domain explored, and for the Li and Be abundances in G-type stars, they fail to explain the Li and Be dips in these two clusters. On the other hand, $^M_{\nu'}R2^{T6.42}_A$ models with the parametric viscosity for the transport of angular momentum, dependent on the stellar mass of the star and adjusted to fit the Li abundances along the entire mass range, predict a Be dip that is not as deep as the one observed, and they lead the 1 M$_{\odot}$ model to rotate too slowly at the age of the two clusters (0.72-0.75~Gyr). 
In Sect.~\ref{sect:multicluster} we evaluate the respective compatibility of the models $_{\nu}R1_A^{T6.42}$ and $^M_{\nu'}R2^{T6.42}_A$ with the surface Li and rotation rates in a sample of open clusters, before discussing in
Sect.~\ref{sect:Discussion} the impact of the different assumptions for the transport of angular momentum on the internal rotation profiles of the models in the light of asteroseismic constraints. In Sect.\ 7, we provide a summary of the successes and difficulties of the different assumptions, and we discuss their meaning in terms of the uncertainties of the different processes involved.

\section{Model predictions for Li and surface rotation - Comparison to open clusters of different ages and metallicities}
\label{sect:multicluster}

In this section, we compare the predictions of models $_{\nu}R1_A^{T6.42}$ and $^M_{\nu'}R2^{T6.42}_A$ to rotation rates and surface Li abundances for a sample of open clusters with different ages and metallicities (see Table~\ref{tab:OCinfos1}). Be cannot be used here, as there are not enough available data for clusters other than the Hyades and Praesepe. 
In the $_{\nu}R1^{T6.42}_A$ case, we computed models for the actual [Fe/H] of the individual clusters and for three initial rotation rates (slow, median, and fast), while in the $^M_{\nu'}R2^{T6.42}_A$ case, we used the models at the metallicity of the Hyades and Praesepe ([Fe/H] = +0.15 dex), and for median rotation discussed previously. 
\begin{figure*}[t]
         \center
         \includegraphics [width=150mm]{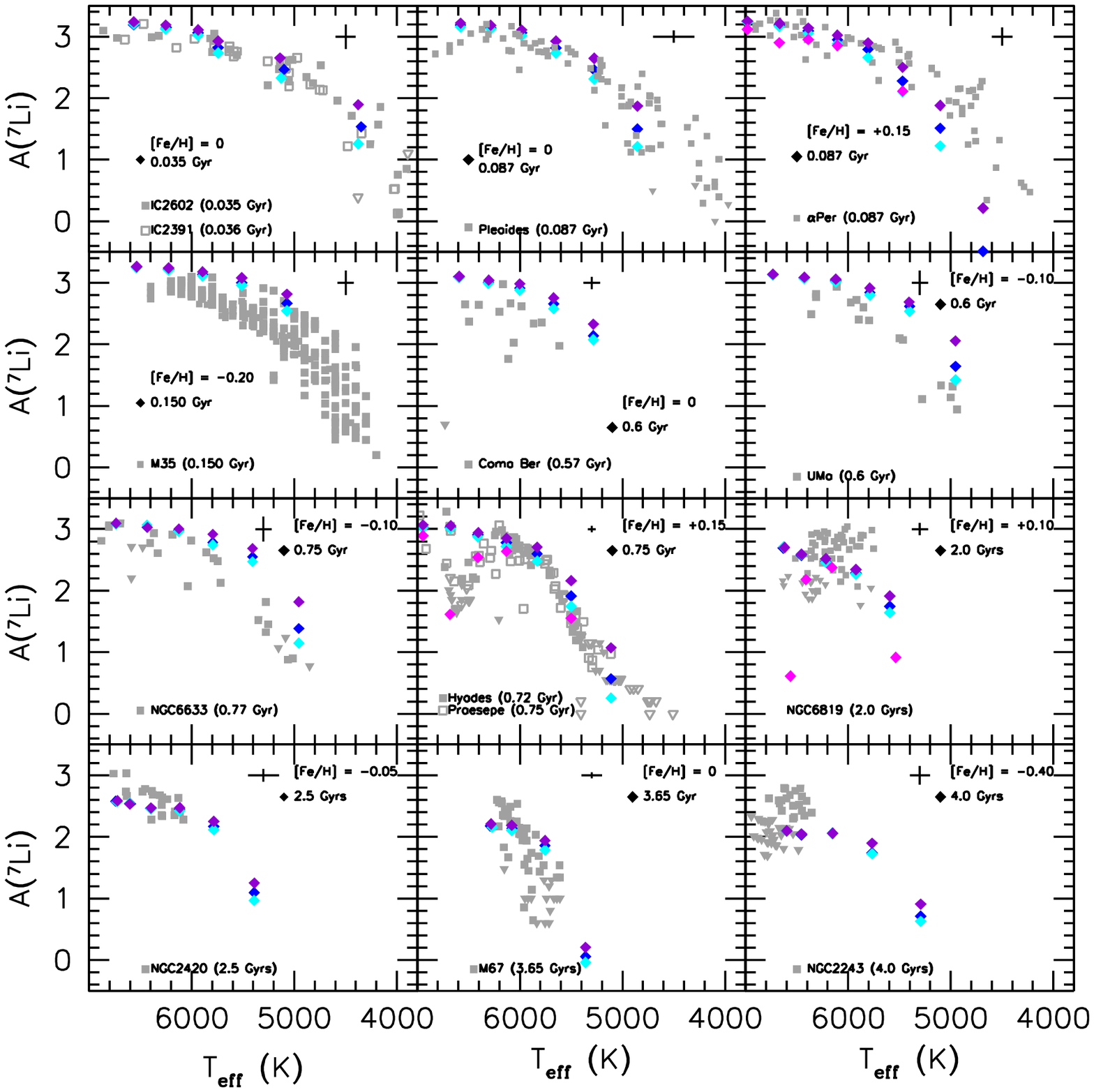}
         \caption{Li surface abundance versus T$_\mathrm{eff}$ in open clusters of different ages and [Fe/H] (see Table~\ref{tab:OCinfos1} for references; grey squares and triangles are abundance determinations and upper limits, respectively). The typical observation errorbars from the original papers are indicated by a cross in each panel. Coloured diamonds are the predictions of the models at the age and metallicity of the corresponding clusters. The $_{\nu}R1^{T6.42}_A$ models are shown for three initial velocities (slow, median, and fast are the cyan, blue, and violet diamonds, respectively), and models $^M_{\nu'}R2^{T6.42}_A$ for median rotation only (magenta diamonds). Models from left to right in each panel correspond to masses between 1.5 M$_\odot$ (warmer) and 0.8 M$_\odot$ (cooler).  
         }
         \label{fig:LiTeffFeH0}
 \end{figure*}
The values of A(Li), T$_{\rm eff}$, and P$_{\rm rot}$ predicted by the $^M_{\nu}R1_A^{T6.42}$ model are given in Appendix~\ref{AnnexeA} for the different masses and cluster ages.

\subsection{P$_{\rm rot}$ versus mass}
\label{sect:protmass}

In Fig.~\ref{fig:Protmass} we compare the predicted surface rotation periods of the $_{\nu}R1_A^{T6.42}$ models with relevant [Fe/H] to observational data for a sample of 12 open clusters at different ages. We added the $^M_{\nu'}R2_A^{T6.42}$ model predictions at the [Fe/H] of the Hyades for $\alpha$Per, Hyades, Praesepe, and NGC 6819. The masses of the individual cluster stars were determined by isochrone fitting in the original papers quoted in Table~\ref{tab:OCinfos1} for the P$_{\rm rot}$. \citet{2021arXiv210101183G} have studied the difference in mass estimates that can be derived from using different families of isochrones and they show that it is quite small ($\Delta M \approx 0.05 {\rm M}_\odot$). 

We confirm the conclusions reached in previous studies \citep[e.g.][]{1997ApJ...480..303K,1998A&A...333..629A,2009IAUS..258..363I,2016A&A...587A.105A,2019A&A...631A..77A}: the evolution of the rotation period can be divided in several phases that our models succeed to predict. Firstly, the large dispersion at very young ages for all masses (NGC 2362 to IC 2391, first row of Fig.~\ref{fig:Protmass}) results in our models from the assumption of constant surface angular velocity as long as disc-locking is efficient. Secondly, the progressive slow down of first, slow and median, and finally fast rotators of decreasing mass in clusters of increasing age (from $\alpha$ Per to M 37, second row of Fig.~\ref{fig:Protmass}) is very well reproduced by our models as the result of the secular evolution of the radius along the PMS and the efficient wind braking on the early MS. 
Finally, for clusters older than Praesepe ($\approx$ 750 Myrs), we observe the convergence of the three families of rotators into one single sequence recovering the observational law by \citet[][]{1972ApJ...171..565S}. In addition, the sequence presents a change in slope around M $\approx$ 1.2 M$_\odot$ in the P$_{\rm rot}$ versus mass diagram (last row of Fig.~\ref{fig:Protmass}), resulting from the magnetic braking by the stellar winds. This change in slope corresponds to the Kraft break and to the mass domain where the convective envelope becomes very thin.
In the Hyades and Praesepe, we also show the predictions of the model $^M_{\nu'}R2_A^{T6.42}$ introduced to fit the cool edge of the Li dip (see Sect.~\ref{subsect:hyades}). The predicted rotation periods for models more massive than 1.2 M$_\odot$ are fully compatible with observations, while the 1 M$_\odot$ median rotator model is far too slow, as already discussed in Sect.~\ref{subsect:hyades}.
As in \citet{2019A&A...631A..77A}, who use the same prescriptions for horizontal and vertical turbulence and for the braking law as in model $_{\nu}R1_A^{T6.42}$ (with a parameter K=7.0$\times 10^{30}$ erg instead of 7.5$\times 10^{30}$ erg) but who do not include the parametric viscosity for the internal transport of angular momentum, our model predictions for the older clusters present a larger negative slope on the lower mass side, with the lower mass models spinning slower than the observed stars. For the more massive stars in NGC 6819, the extraction of angular momentum, which is directly linked to the depth of the convective envelope, becomes inefficient in the models, which spin faster than the observed stars.
Finally, for the Pleiades we obtain a poor agreement with observations for the lowest stellar masses (0.8 and 0.9 M$_\odot$) compared to what we get for all the other clusters and to what was obtained in \citet{2019A&A...631A..77A}. This is essentially due to the difference in age adopted for this cluster in both studies (125~Myrs from \cite{2004ApJ...614..386B} in \cite{2019A&A...631A..77A} instead of 87~Myrs as assumed here), as shown in Fig.~\ref{fig:Protmass} where the orange diamonds are the predictions of our models at 125~Myrs, which better account for the data.

To summarise, the comparison between the predictions of our $_{\nu}R1^{T6.42}_A$ and $_{\nu'}^{M}R2^{T6.42}_A$, and the observational data for the rotation period is rather satisfactory, although some peculiar features remain difficult to reproduce without further adjustments. This work confirms the global theoretical behaviour obtained by \citet[][]{2016A&A...587A.105A,2019A&A...631A..77A} despite the fact that we included an additional diffusive source of transport of angular momentum in the present study.

\subsection{Lithium - T$_{\rm eff}$}

In Fig.~\ref{fig:LiTeffFeH0}, we compare the predictions of models $_{\nu}R1^{T6.42}_A$ for the surface Li abundances to data within the cluster sample (see Sect.~\ref{sec:obs} and Table~\ref{tab:OCinfos1}), and we report the predictions of models $_{\nu'}^{M}R2^{T6.42}_A$ discussed in Sect.~\ref{sect:predictions} for the most metal-rich clusters ($\alpha$Per, Hyades, Praesepe, and NGC 6819). 
Importantly, over the entire mass and metallicity range explored, the faster the initial rotation velocity, the lower the predicted Li depletion. As already discussed in Sect.~\ref{subsect:hyades}, this is due to penetrative convection acting on the PMS, which we simulated with the \citet{2019ApJ...874...83A} prescription.

The Li-T$_{\rm eff}$ relation for G-type stars is well reproduced by models $_{\nu}R1^{T6.42}_A$ at all ages from the youngest (IC~2602 and IC~2391) to the oldest cluster (M67) with solar or super-solar metallicities. For the young clusters with subsolar metallicity (M35, NGC~6633, and UMa), the cooler models, corresponding to lower mass stars (see Table in Appendix A), however predict larger A(Li) values than observed. 
In these models, the convective envelope is thinner due to the lower metallicity and the penetrative convection, which is the main process responsible for the Li depletion on the PMS and it is less efficient than at higher metallicity. This can be directly seen when comparing the panels showing NGC~6633 and UMa clusters as well as the Hyades and Praesepe clusters, which have similar ages but different metallicities. This suggests a stronger metallicity dependence of the Li depletion in cool stars compared to that predicted. Data for the cooler stars in the most metal-poor and oldest cluster (NGC~2243) would be required to confirm this trend at older ages. 

\begin{figure}[t]
         \center
         \includegraphics [width=90mm]{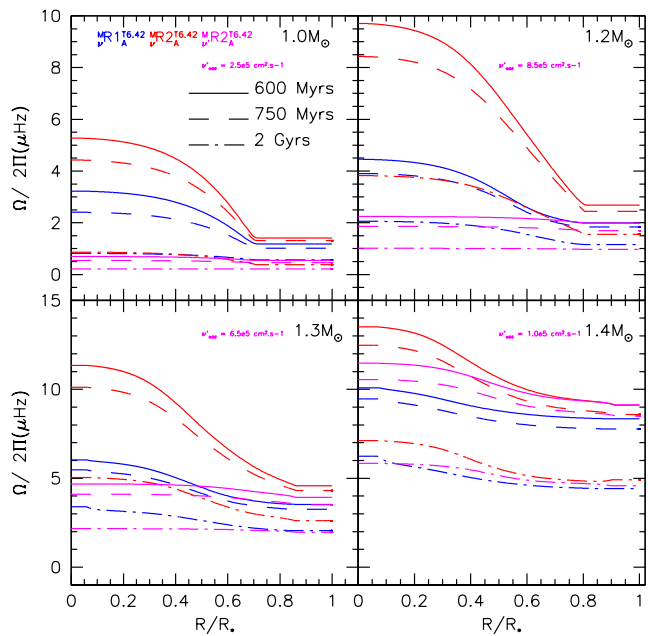}
         \caption{Angular velocity profiles versus the radius within the star for models $^M_{\nu}R1^{T6.42}_A$ (blue), $^M_{\nu}R2^{T6.42}_A$ (red), and $^M_{\nu'}R2^{T6.42}_A$ (magenta) at [Fe/H]=+0.15 and for four masses: 1.0, 1.2, 1.3, and 1.4$M_{\odot}$ (the value of $\nu'_{\rm add}$ used is indicated for each mass). The solid, dashed, and dot-dashed lines refer to three different ages (0.6, 0.75, and 2~Gyrs, respectively).}
         \label{fig:profomega}
 \end{figure}

We see in Fig.~\ref{fig:LiTeffFeH0} that the Li dip is not present in the youngest clusters, and that its depth increases with time along the MS \citep[see also][]{Wallerstein1965,1986ApJ...302L..49B,1986ApJ...309L..17H,1993AJ....106.1080S,Balachandran1995,2004ApJ...614L..65S,2009AJ....138.1171A,2012AJ....144..137C,2016ApJ...830...49B}.
As already discussed in Sect.~\ref{subsect:hyades}, model $_{\nu}R1^{T6.42}_A$ predicts too little Li depletion in the dip, while $_{\nu'}^{M}R2^{T6.42}_A$ models (computed only for [Fe/H]=+0.15~dex) reproduce both the depth and the shape of the Li dip in clusters with super-solar metallicities and with ages between 720~Myrs (Hyades) and 2~Gyrs (NGC 6819). 
For the old low metallicity NGC 2243 cluster, the cold edge of the Li dip is shifted towards slightly higher T$_{\rm eff}$ compared to the younger higher metallicity clusters \citep{2012AJ....144..137C,2013A&A...552A.136F}.
Since models $_{\nu}R1^{T6.42}_A$ lie below the observational points on the cold edge of the Li dip, we expect that model $_{\nu'}^{M}R2^{T6.42}_A$, assuming a similar dependence between ${\nu'_{\rm add}}$ and T$_{\rm eff}$ as assumed to fit the Hyades and Praesepe, would  predict an even stronger Li depletion and would be incompatible with the data of NGC 2243. 

We thus reach similar conclusions as in Sect.~\ref{subsect:hyades}, for a relatively large range of metallicities and ages. Model $_{\nu}R1^{T6.42}_A$ initially developed to reproduce surface and internal constraints for solar-type stars nicely explains the general trend of the data for both the rotation and Li behaviours over the mass and the metallicity ranges probed in this study, except for the Li dip. On the other hand, model $^M_{\nu'}R2^{T6.42}_A$ with turbulent viscosity for the transport of angular momentum parametrised to fit the dip in the Hyades and the Pleiades predicts rotation rates that are too slow for the lower mass models. Importantly, and for the first time, we predict that over the entire mass and metallicity range explored here, penetrative convection that is efficient at the early stages of the PMS leads to larger Li depletion in the slower rotators. This anti-correlation between the rotation rate and the Li abundance, which builds early on the PMS, remains constant over time.

 \begin{figure}[t]
         \center
         \includegraphics [width=90mm]{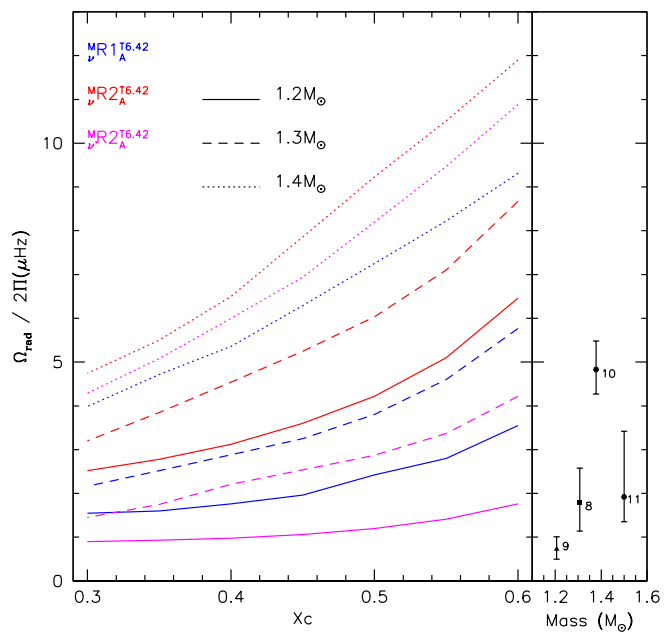}
         \caption{(Left) Mean angular velocity within the stellar layers in the radiative zone as a function of the  remaining central hydrogen mass fraction $X_c$ (time increases from right to left) for the same models as in Fig.~\ref{fig:profomega}. Masses 1.2, 1.3, and 1.4$M_{\odot}$ are represented by full, dashed, and dotted lines respectively. (Right) Stars observed by \citet[][]{2015MNRAS.452.2654B} as indexed in the original paper (8: $M=1.307\pm0.027$ $M_{\odot}$, 9: $M=1.206\pm0.077$ $M_{\odot}$, 10: $M=1.378\pm0.028$ $M_{\odot}$, and 11: $1.500\pm0.050$ $M_{\odot}$). 
         }
         \label{fig:benomar}
 \end{figure}

\section{Model predictions for internal rotation and comparison to asteroseismic constraints} 
\label{sect:Discussion}

In this section, we present the predictions of models $^M_{\nu}R1_A^{T6.42}$, $^M_{\nu}R2_A^{T6.42}$, and $^M_{\nu'}R2^{T6.42}_A$ with [Fe/H] = +0.15 dex (characteristic of the Hyades, Praesepe, and NGC 6819 clusters) in terms of internal rotation and tentatively compare them to the only available asteroseismic estimates of internal rotation rates in MS field stars of a similar metallicity from \cite{2015MNRAS.452.2654B}. 

In Fig.~\ref{fig:profomega}, we present the internal rotation profiles for the 1.0, 1.2, 1.3, and 1.4~M$_\odot$ models at three different ages on the MS. 
For G-type stars, the $^M_{\nu}R1_A^{T6.42}$ model that reproduces all surface observational constraints (see previous sections) predicts a small radial differential rotation that decreases with time and is compatible with helioseismic data as already shown in Paper I. For the more massive models corresponding to the Li-dip stars in the Hyades, Praesepe, and NGC 6819, this model predicts the same behaviour.
The $^M_{\nu}R2_A^{T6.42}$ model predicts larger radial differential rotation for all masses and ages than the $^M_{\nu}R1_A^{T6.42}$ model, which is in agreement with the associated strong Li depletion discussed in previous sections. This results from the much higher efficiency of the vertical turbulent shear D$_v$ when using the \cite{1997A&A...317..749T} prescription (see also Paper I).
For the $^M_{\nu'}R2^{T6.42}_A$ model, which introduces an enhanced parametric turbulent transport of angular momentum to fit the cold edge of the Li dip, the radial differential rotation is strongly reduced at all ages, especially in the 1.2 M$_\odot$ model with the larger value of $\nu'_{\rm add}$, which presents an almost flat rotation profile at all ages displayed on the figure. Concerning the case of the 1.4 M$_\odot$ model, although the angular momentum contrast between the surface and the core is of the same order in models $^M_{\nu}R1^{T6.42}_A$ and $^M_{\nu'}R2_A^{T6.42}$, the angular velocity gradient near the base of the convective envelope is much larger in the later model (see change in slope in magenta profiles around r = 0.9 R$_\star$ in Fig.~\ref{fig:profomega}), thus leading to an enhanced shear and associated turbulent transport in the exact region where the connection between the convective envelope and Li burning zone is made. This leads to the strong Li depletion previously discussed in this model (see Sect.~\ref{subsect:hyades}). On the contrary, the shear is near to null at the base of the convective envelope in model $^M_{\nu}R1^{T6.42}_A$, leading to almost no Li depletion.

 Fig.~\ref{fig:benomar} shows the mean angular velocity in the radiative interior (i.e. excluding the convective core if present\footnote{The expression used is similar to Eq.~(18) from Paper I, but starting the integration at the outer edge of the convective core when present: $\overline{\Omega_{\rm rad}} = \int_{M_{\rm TCC}}^{M_{\rm BCE}} r^2 \Omega \, dm / \int_{M_{\rm TCC}}^{M_{\rm BCE}} r^2 dm$, with $M_{\rm TCC}$ being the mass coordinate of the top of the convective core and $M_{\rm BCE}$ being the mass coordinate at the base of the convective envelope.}, which is consistent with asteroseismic data that do not take it into account) as a function of the central hydrogen mass fraction in the 1.2, 1.3, and 1.4~M$_{\odot}$ models at [Fe/H]=+0.15~dex. We compare this to a selected sub-sample of four field stars with metallicities close to that of the Hyades for which the average rotation in the radiative interior was estimated from asteroseismology by \citet[][]{2015MNRAS.452.2654B}. Our predictions are of the same order of magnitude as the results from \citet[][]{2015MNRAS.452.2654B}, but the predicted rotation is faster than the values they inferred for the selected targets. 
 Models $^M_{\nu'}R2^{T6.42}_A$ lead to a better agreement with the values derived from asteroseismic data than models $^M_{\nu}R2^{T6.42}_A$. A more precise comparison with the results in \cite{2015MNRAS.452.2654B} is prevented by the use of MESA stellar models in their study including very different input physics and settings. In particular, core hydrogen mass fractions $X_c$ cannot be compared.

Additional data for different stellar masses would be required to better discriminate between the different prescriptions for the transport of angular momentum, especially for cool stars. Nevertheless, this comparison supports the need for strong transport of angular momentum on the MS for stars in the Li-dip region, as obtained when increasing the strength of $\nu_{\rm add}$ as assumed in models $^M_{\nu'}R2^{T6.42}_A$.

\section{Summary and discussion} 
\label{sect:CONCLUSION}

The need for additional transport processes beyond atomic diffusion and so-called Type-I rotation-induced processes (turbulent shear and meridional circulation) for the transport of chemicals and angular momentum has long been reported (see references in Sect.~\ref{section:introduction}). This work follows Paper I analysing the impact of several processes, using state-of-the-art prescriptions for specific mechanisms (rotation-induced turbulence and penetrative convection in particular), as well as parametric prescriptions for others (parametric turbulence for chemicals and parametric viscosity for angular momentum) that were calibrated to reproduce the evolution of the surface Li abundance and rotation rates as well as the internal angular velocity profiles in the Sun and solar-type stars. The aim of the present paper is to study the impact of these processes (using the solar-type calibrations) on F- and G-type MS stars at different metallicities and to test them against data in Galactic open clusters over a large range in ages. Although the adopted ages come from different references and are not fully consistent, our conclusions are robust given the uncertainties on their estimate. Our predictions support, however, the Pleiades to be older than assumed in this work, which is in agreement with other age determinations for this cluster. 

The so-called $_{\nu}R1^{T6.42-6.5}_A$ models calibrated in Paper I include atomic diffusion, rotation-induced processes (meridional circulation and turbulence), penetrative convection, parametric turbulence for the transport of chemicals, and parametric viscosity for the transport of angular momentum.
We have also presented $^M_{\nu'}R2^{T6.42}_A$ models which differ by the assumed prescriptions for horizontal and vertical shear-induced turbulence and were computed with different values for the parametric viscosity.
We have analysed the agreement between the theoretical predictions of these two sets of models and observational data for Li, Be, and surface rotation rates available in a sample of open clusters of different ages and metallicities. We also compared the predicted internal rotation profiles with asteroseismic constraints in field MS stars with [Fe/H] close to that of the Hyades.

Both $_{\nu}R1^{T6.42-4.5}_A$ and $^M_{\nu'}R2^{T6.42}_A$ explain the main general trends observed between Li depletion and stellar mass, age, rotation, and metallicity covered in this study. Thanks to the prescription we used for penetrative convection \citep{2019ApJ...874...83A}, an anti-correlation between Li depletion and surface rotation build up early on the PMS, and this remains throughout the evolution on the MS.
Theoretical surface Li abundance and T$_{\rm eff}$ were correlated as observed in cluster stars, with cooler and lower mass stars being more Li depleted than hotter and more massive ones (except in the Li dip when this feature is present).  
Our models also recover the anti-correlation between metallicity and Li depletion efficiency that is observed in open clusters stars. Metal-poor stars are less Li-depleted than their metal-rich counterparts, as expected from the metallicity dependence of the location of the base of the convective envelope that affects the PMS Li depletion.
However, Li depletion in our metal-poor models is more modest than in open cluster stars. 

Model $_{\nu}R1^{T6.42-6.5}_A$ succeeds to reproduce the Li, Be, and rotation rates observations in G-type stars at all ages, and in F-type stars in the young clusters where the Li dip has not started to form yet. This is achieved thanks to the combined effects of penetrative convection, rotation-induced processes, and parametric turbulence, whose efficiencies vary with both the stellar mass and the metallicity.
The impact of these mechanisms on Li and Be depletion becomes minute in the more massive stars that have a very thin convective envelope, more distant from the Li and Be burning layers. As a result, model $_{\nu}R1^{T6.42-6.5}_A$ is not able to reproduce the Li and the Be dips centred around $\sim$6'600 K that appear later on the MS, although it reproduces the rotation rates over the entire mass range covered in this study. On the other hand, model $^M_{\nu}R2^{T6.42}_A$ computed with the same value of the parametric viscosity, but with different prescriptions for horizontal and vertical turbulence ($D_h$ and $D_v$) that lead to more differential rotation (hence more efficient rotation-induced mixing) in the stellar interior, predicts too much Li depletion on the cool side of the dip, which is as expected from previous studies. 

Building on the assumption that an additional transport process such as internal gravity waves would be the main driver for the transport of angular momentum in stellar interiors on the cold edge of the Li dip, we introduced a mass-dependent viscosity $\nu'_{\rm add}$ with a maximum efficiency at $\approx$ 5'900~K as predicted in \citet{2003A&A...405.1025T} and \citet{2004A&A...418.1051T} for the IGW excitation by the convective envelope and their luminosity. 
Although the assumption we make of a parametric viscosity $\nu^{(')}_{\rm add}$ being uniform within the radiative interior and constant with time is a very crude parametrisation of a transport of angular momentum within stars, the $^M_{\nu'}R2^{T6.42}_A$ model predicts a good agreement with Li abundances on the cold edge of the dip in open clusters of different ages at the metallicity of the Hyades (this was not tested for other metallicities). This is due to the flattening of the angular velocity gradient within the stellar interior, and in particular between the base of the convective envelope and the Li burning depth, implied by higher values of the parametric viscosity. A further improvement to our model now requires the full treatment of the transport of angular momentum by IGWs as in \citet{2005Sci...309.2189C}, also taking into account  recent developments on the generation and the behaviour of IGWs \citep[e.g.][]{2016A&A...588A.122P,2020ApJ...903...90A,2020MNRAS.497.4231R}. The Tayler-Spruit dynamo process that was updated in recent studies \citep[e.g.][]{2019MNRAS.485.3661F,2019A&A...626L...1E,2019A&A...621A..66E,2019A&A...631L...6E} is also a candidate for additional transport of angular momentum and should also be tested regarding the results of the present work.
The Be depletion obtained with model $^M_{\nu'}R2^{T6.42}_A$ is on the other hand too modest to explain the depth of the Be dip observed in a couple of clusters. 
The acquisition of more Be observational data in open clusters of different ages and metallicities, in addition to those in field stars \citep[e.g.][]{2012ApJ...746...47D}, would be an important key to better constrain models.

Importantly, the transport of angular momentum simulated with the parametric viscosity values adopted in our different models is much more efficient than that driven by meridional circulation and turbulence. Both the $_{\nu}R1^{T6.42}_A$ and the $^M_{\nu'}R2^{T6.42}_A$ models predict mean angular rotation velocity in the radiative interior that are compatible with the values inferred from asteroseismology in field MS with the same metallicity as the Hyades and the Pleiades. Additional asteroseismic data to probe the internal rotation of young MS stars over a large metallicity range would be valuable.

Reflecting on the modelling of turbulent shear, we have to conclude that none of the combinations for the prescriptions  of the horizontal and vertical turbulence ($D_h$ and $D_v$) we used is able to reconcile all the surface constraints used in this study.
Model $_{\nu}R1^{T6.42-6.5}_A$, tailored for solar-type stars, includes the \citet{2018A&A...620A..22M} prescription for the horizontal shear diffusivity $D_h$, which is the only prescription including the contributions of both the horizontal and vertical shear on the horizontal turbulent transport, and the \citet{1992A&A...265..115Z} prescription for the vertical shear diffusivity $D_v$, which does not take into account the effects of thermal and molecular diffusivity on the vertical turbulent transport, but it is the only one that has been validated by direct numerical simulations \citep[][]{2013A&A...551L...3P,2017ApJ...837..133G}. This model thus includes the apparently most robust available prescriptions, but it cannot explain the Li dip even when combined with additional sources for turbulent transport such as D$_{\rm T_0}$ and  $\nu_{\rm add}$. 

On the other hand, the use of the \citet{1992A&A...265..115Z} prescription for the horizontal shear diffusivity ($D_h$) and that of \citet{1997A&A...317..749T}, which includes the effect of thermal and molecular diffusion for the vertical shear diffusivity ($D_v$)  in model $^M_{\nu'}R2^{T6.42}_A$, requires higher values for the parametric viscosity for the internal transport of angular momentum, with a mass-dependent efficiency as expected from IGW. The choice of these prescriptions thus impacts our ability to constrain the other transport processes \citep[see][]{2013LNP...865....3M,2016A&A...587A.105A} as additionally illustrated by the predictions obtained in this work and for instance by \citet{2020A&A...643A.164S} for the specific case of NGC~2420.  
Insight and validation of the different available prescriptions for the vertical turbulent shear transport from hydrodynamicists and multi-dimensional numerical simulations, similar to the ongoing work on the horizontal turbulent shear \citep{2020A&A...635A.133P,2021A&A...646A..64P,2021A&A...649A..62P}, are now mandatory in order to overcome this impasse.

Finally, two additional promising leads to understand the formation of the Li and Be dips, in particular, should be explored, which were neglected here.
First, we do not consider radiative accelerations on heavy elements. %
Their effects are, however, non-negligible for stars with an effective temperature higher than $\approx$ 6'800~K, which corresponds to $\approx$ 1.4~M$_{\odot}$ at solar metallicity, close to the hot edge of the Li dip \citep[e.g.][]{1993ApJ...416..312R,2018A&A...618A..10D,2020A&A...633A..23D}. Their impact on the stellar opacity could  partly modify some of our conclusions concerning the most massive models. Second, the process sustaining the parametric turbulence used in this work and others in the literature has yet to be identified. In particular, even if the tachocline mixing \citep[][]{1992A&A...265..106S,1999ApJ...525.1032B,2020ASSP...57..207G} was shown to not be adapted in solar-type stars (Paper I), this process could play a role in the building of the Li and Be dips because of its dependence at the latitudinal differential rotation that is predicted to scale inversely with rotation for F-type stars \citep[][]{2012ApJ...756..169A} compared to G-type stars.

\begin{acknowledgements}
 We thank J.W.~Ferguson for providing us with low-temperature opacity tables adapted to the solar mixture we adopt in this work, E.~Semenova for providing us observational data for NGC~2420 open cluster, J.D.~Cummings and collaborators for providing the log~g values for their Hyades and Praesepe sample, L.~Amard and P.~Eggenberger for fruitful discussions, and the anonymous referee for constructive comments on the manuscript. 
    This work was supported by the Swiss National Science Foundation (Project 200020-192039 PI C.C.), by CNES (funding of AP at LUPM) focussed on the PLATO mission, and by the Agence Nationale de la Recherche grant POPSYCLE number ANR-19-CE31-0022. This research has made use of NASA's Astrophysics Data System Bibliographic Services.  
\end{acknowledgements}

%
%

\bibliographystyle{aa}
\bibliography{references}

\clearpage
\onecolumn

\appendix
\section{Model predictions}
\label{AnnexeA}

\begin{longtable}{c c c c c}
\caption{$^M_{\nu}R1^{T6.42}_A$ model predictions for the different clusters with lithium abundances presented in Table~\ref{tab:OCinfos1} and in Figs.~\ref{fig:Protmass} and \ref{fig:LiTeffFeH0}. }\\
\hline\hline
Cluster & Mass (M$_\odot$) & T$_{\rm eff}$ (K) &A(Li) & P$_{\rm rot}$ (d)\\ \hline
 \endfirsthead
\caption{continued.}\\        
\hline\hline
 Cluster & Mass (M$_\odot$) & T$_{\rm eff}$ (K) &A(Li) & P$_{\rm rot}$ (d)\ \\ \hline
\endhead
\hline
\endfoot
 & 0.8 & 4347 & 1.54 & 1.6 \\
   &      0.9 & 5099 & 2.47 & 1.5 \\
IC 2602 / IC 2391    &     1.0 & 5742 & 2.82 & 1.2 \\
     &    1.1 & 5937 & 3.06 & 1.2\\
      &   1.2 & 6247 & 3.15 & 1.3 \\
       &  1.3 & 6564 & 3.19 & 1.1 \\
       \hline
  &  0.8 & 4855 & 1.50 & 1.3 \\
    &     0.9 & 5275 & 2.46 & 1.5 \\
Pleaides    &     1.0 & 5649 & 2.81 & 1.8 \\
    &     1.1 & 5982 & 3.06 & 2.4 \\
    &     1.2 & 6284 & 3.14 & 1.9 \\
    &     1.3 & 6578 & 3.19 & 0.9 \\
         \hline  
 & 0.8 & 4685 & -0.49 & 1.3 \\
  &       0.9 & 5103 & 1.51 & 1.5 \\
  &       1.0 & 5470 & 2.28 & 1.8 \\
$\alpha$Per  &       1.1 & 5806 & 2.79 & 2.4 \\
  &       1.2 & 6104 & 2.95 & 1.9 \\
  &       1.3 & 6388 & 3.09 & 0.9 \\
  &      1.4 & 6670 & 3.19 & 2.3 \\
         \hline
  & 0.8 & 5072 & 2.67 & 2.3 \\
   &      0.9 & 5512 & 3.00 & 2.6 \\
 M 35&         1.0 & 5894 & 3.13 & 3.4 \\
&         1.1 & 6225 & 3.22 & 3.5 \\
&         1.2 & 6538 & 3.25 & 2.4\\ 
         \hline
 & 0.9 & 5284 & 2.13 & 10.5 \\
  &       1.0 & 5671 & 2.66 & 8.5 \\
Coma Ber  &       1.1 & 6001 & 2.92 & 6.6 \\
  &       1.2 & 6305 & 3.01 & 4.3 \\
  &       1.3 & 6597 & 3.10 & 1.1 \\
         \hline
  &       0.8 & 4953 & 1.64 & 11.8 \\
  &       0.9 & 5402 & 2.62 & 9.3 \\
  UMa &       1.0 & 5789 & 2.84 & 8.0 \\
      &   1.1 & 6121 & 3.03 & 5.7 \\
      &   1.2 & 6429 & 3.08 & 3.2 \\
      &   1.3 & 6733 & 3.14 & 1.3 \\
         \hline
 &   0.8 & 4665 & -2.11 & 12.4 \\
 &        0.9 & 5108 & 0.85 & 10.5 \\
 &        1.0 & 5497 & 1.99 & 8.5 \\
  Hyades &       1.1 & 5829 & 2.65 & 6.6 \\
    &     1.2 & 6127 & 2.83 & 4.3\\
    &     1.3 & 6407 & 2.96 & 3.3 \\
    &     1.4 & 6679 & 3.07 & 1.4 \\
    &     1.5 & 6964 & 3.08 & 0.8 \\
         \hline 
&         0.8 & 4955 & 1.38 & 14.3\\
&         0.9 & 5406 & 2.55 & 10.5\\
&         1.0 & 5793 & 2.79 & 9.0 \\
NGC 6633    &     1.1 & 6126 & 2.98 & 6.3 \\
    &     1.2 & 6434 & 3.03 & 3.5 \\
    &     1.3 & 6737 & 3.09 & 1.4 \\
         \hline    
 &        0.8 & 4667 & -2.73 & 15.2 \\
 &        0.9 & 5112 & 0.57 & 12.1\\
 &        1.0 & 5501 & 1.91 & 9.7\\
 Praesepe &        1.1 & 5833 & 2.60 & 7.2 \\
    &     1.2 & 6132 & 2.78 & 4.7 \\
    &     1.3 & 6410 & 2.91 & 3.6 \\
    &     1.4 & 6677 & 3.03 & 0.9 \\
    &     1.5 & 6944 & 3.03 & 0.8 \\
         \hline 
  &      0.9 & 5198 & -1.32 & 24.3 \\
NGC 6819  &         1.0 & 5594 & 1.74 & 19.1\\
&         1.1 & 5925 & 2.29 & 13.1 \\
    &     1.2 & 6218 & 2.47 & 8.6 \\
NGC 6819     &     1.3 & 6457 & 2.57 & 4.5 \\
    &     1.4 & 6634 & 2.70 & 1.8 \\
         \hline
   &      0.9 & 5391 & 1.10 & 29.1 \\
&         1.0 & 5785 & 2.17 & 18.9 \\
NGC 2420&         1.1 & 6118 & 2.45 & 11.9 \\
        & 1.2 & 6398 & 2.46 & 6.9 \\
        & 1.3 & 6598 & 2.54 & 3.0 \\
        & 1.4 & 6741 & 2.58 & 1.2 \\
         \hline
     &    0.9 & 5369 & -0.24 & 49.5\\
     &    1.0 & 5765 & 1.78 & 28.7 \\
     M67 &    1.1 & 6083 & 2.05 & 17.5  \\
    &     1.2 & 6240 & 2.10 & 11.7 \\
         \hline 
   &      0.8 & 5293 & 0.71 & 48.5 \\
&         0.9 & 5766 & 1.74 & 26.6 \\
 NGC 2243 &        1.0 & 6152 & 2.06 & 15.3 \\
        & 1.1 & 6455 & 2.04 & 7.2 \\ \hline
\end{longtable}

\end{document}